\documentclass[twocolumn,pre,superscriptad16ess,aps,nofootinbib]{revtex4}

\usepackage{epsfig}
\usepackage{amsmath}
\usepackage{graphicx}
\usepackage{amsmath}
\usepackage{amsthm}
\usepackage{multirow}
\usepackage{relsize}
\usepackage{amssymb}
\usepackage{bm}
\usepackage{textcomp}

\usepackage{epstopdf}

\DeclareMathAlphabet{\mathitbf}{T1}{cmr}{bx}{it}

\begin{document}

\title{Critical behavior of the three-dimensional random-anisotropy Heisenberg model}

\author{J.J. Ruiz-Lorenzo} \affiliation{ Departamento de F\'{\i}sica
  and Instituto de Computaci\'{o}n Cient\'{\i}fica Avanzada (ICCAEx),
  Universidad de Extremadura, 06071 Badajoz, Spain.}  \affiliation{
  Instituto de Biocomputaci\'{o}n y F\'{\i}sica de Sistemas Complejos
  (BIFI), 50018 Zaragoza, Spain.  }

\author{M. Dudka}
\affiliation{Institute  for Condensed   Matter  Physics,  National  Acad.  Sci.   of
Ukraine, UA--79011 Lviv, Ukraine}\affiliation{${\mathbb L}^4$ Collaboration \& Doctoral College for the Statistical Physics of Complex Systems, Leipzig-Lorraine-Lviv-Coventry, Europe}

\author{Yu. Holovatch}
\affiliation{Institute  for Condensed   Matter  Physics,  National  Acad.  Sci.   of
Ukraine, UA--79011 Lviv, Ukraine}\affiliation{${\mathbb L}^4$ Collaboration \& Doctoral College for the Statistical Physics of Complex Systems, Leipzig-Lorraine-Lviv-Coventry, Europe}\affiliation{Centre for Fluid and Complex Systems, Coventry University, Coventry, CV1 5FB, United Kingdom}\affiliation{Complexity Science Hub Vienna, 1080 Vienna, Austria}

\date{\today}

\begin{abstract}

We have studied the critical properties of the three-dimensional
random anisotropy Heisenberg model by means of numerical simulations
using the Parallel Tempering method. We have simulated the model with
two different disorder distributions, cubic and isotropic ones, with
two different {anisotropy} strengths for each disorder class.
For the case of the anisotropic disorder, we have found
evidences of universality by finding critical exponents and universal
dimensionless ratios independent of the strength of the
disorder. In the case of isotropic disorder distribution the
  situation is very involved: we have found two phase transitions
  in the magnetization channel which
  are merging for larger lattices remaining a 
  zero magnetization low-temperature phase. Studying this region using a 
  spin-glass order parameter we have found evidences for a spin-glass phase transition. We have
  estimated effective critical exponents for the spin-glass phase transition
  for the different values of the strength of the isotropic disorder,
  discussing the cross-over regime.

\end{abstract}


\maketitle

\section{Introduction}\label{I}

Structurally disordered systems apart from fundamental interests are also
important for modern technology.  For instance recently developed rare-earth
based magnetic glasses with magnetocaloric effects were considered as good
candidates for magnetic refrigerants~\cite{Luo_company}. These rare-earth
based systems belong to a wide class of disordered materials~\cite{Hertz,Dudka02} known as random-anisotropy magnets.

Magnetic properties of these systems are described by random anisotropy model (RAM), in which  each
spin is subjected to a local anisotropy of random orientation with the Hamiltonian~\cite{Harris73}:
\begin{equation}
{\mathcal H} =  - J \sum_{\left<{\mathitbf r},{\mathitbf r'}\right>}  \mathitbf{S}_{\mathitbf r}\cdot \mathitbf{S}_{\mathitbf r'}
-D\sum_{\mathitbf r} (\hat{\mathitbf{x}}_{\mathitbf r}
\cdot\mathitbf{S}_{\mathitbf r})^2 \,.
\label{origham}
\end{equation}
Here, $\mathitbf{S}_{\mathitbf r}$ is a classical
$m$--component unit vector on the site $\mathitbf{r}$ of a $d$--dimensional
(hyper)cubic lattice, $D>0$ is the strength of the anisotropy,
$\hat{\mathitbf x}_{\mathitbf r}$ is a (quenched) random unit vector pointing in the
direction of the local anisotropy axis.  The interaction $J>0$ is assumed to be ferromagnetic. We
consider the case of short-range interactions, with $\left<\cdot,\cdot\right>$ meaning summation over pairs of nearest neighbors. The strength  of
the disorder is controlled by the ratio $D/{J}$. 
For the ordered crystalline material, the anisotropy has a well defined direction along one of the coordinate axes, 
which in turn may favor certain types of spin in-plane  alignments.
Obviously, the random
orientations are present in the model (see Eq. (\ref{origham})) only for $m>1$. In the
case $m=1$ the random anisotropy term becomes constant and leads only
to a shift of the Ising system free energy.

Experimental data from random anisotropy systems are accessible from
reviews, see for instance Refs.~\cite{Cochraine78,rev}.  The RAM was
also an object of extensive theoretical and numerical studies reviewed
in Refs.~\cite{Goldschmidt_rev,Dudka05}.  Despite the efforts made so
far, the problem of the nature of a low-temperature phase in random
anisotropy magnets remains a most controversial issue.  In particular,
the question of whether the low-temperature phase is long-range
ferromagnetically ordered or is it a spin-glass.  Local anisotropy
prevents fully ferromagnetic state, where all spins align in the same
direction. Therefore the term ``asperomagnetic'' was proposed for
magnetically ordered low-temperature state with nonzero
magnetization, while the term ``speromagnetic'' was coined for the
spin-glass like state with zero magnetization~\cite{Coey78}. Within
phenomenological theory the last state is also called ``correlated
spin-glass''~\cite{ChudnovskySaslowSerota}. Another possible candidate
for the low-temperature phase in RAM is a quasi long-range order
(QLRO), i.e., the low-temperature ordering typical for the
Berezinskii-Kosterlitz-Thouless
transition~\cite{Berezinskii,KosterlitzThouless}, where the pair
correlation function is characterized by a power-law decay with
distance while the magnetization is zero.

Influence of the local anisotropy axis distributions on the critical
behavior of RAM is of particular interest. The studies performed so
far agree on the fact that the form of the distribution is crucial for
the critical behavior of RAM. In the next section we will review
results of previous studies. We will concentrate mainly on the two
mentioned issues: origin of the low-temperature phase and the role of
the local anisotropy axis distribution. We will leave out of scope
several other challenging topics, such as dynamical aspects of phase
transitions in RAM~\cite{dynamics}, effects of long-range correlations
of local anisotropy axes~\cite{Fedorenko2007} or the presence of a
surface~\cite{Fedorenko2012}.

We will do this with a purpose to emphasize the main goal of our
paper: In the current situation it is of primary importance to apply
the state-of-the-art numerical techniques to get a high accuracy
quantitative description of the critical behavior of three-dimensional
Heisenberg model with moderate quenched random anisotropy in order to
confront predictions of the most recent analytic calculations based on
perturbative renormalization group (RG).

To this end, we will use the Parallel Tempering (PT) method and carefully study low-temperature
behavior of three-dimensional RAM at $m=3$ for two different local anisotropy axis distributions,
the discrete and the continuous one, taking different values of the anisotropy strength for each distribution.

The outline of the rest of the paper is as follows.  We review
theoretical and numerical studies of RAM focusing mainly on results
for the three-dimensional case in Section~\ref{II}. Next, we will
describe our numerical simulations (simulating two different
distributions of anisotropy axis) in Section~\ref{III}.  Our results
are displayed in Section~\ref{IV} and finally we will discus them and
state the conclusions in the last Section~\ref{V}. Technical details
of numerical simulations, and a description of the analysis methods and
the numerical study of the three-dimensional $O(3)$ model, the
two-dimensional $O(2)$ model (as a proxy of a model with QLRO) and the
three-dimensional Edwards-Anderson model (a classic spin glass) are
given in seven appendixes.

\section{Review \label{II}}

\subsection{Theoretical results}

First, we will describe the theoretical results regarding the isotropic disorder distribution.

The earliest theoretical investigations of the RAM were performed
within the mean-field framework.  Ferromagnetism was
predicted~\cite{Harris73,mean} but the possibility of a spin-glass
(SG) phase~\cite{HarrisZobin77} was not excluded.  Exact solution of
the infinite-range interaction limit of the RAM within the mean-field
approach indicates a second-order phase transition to ferromagnetic
(FM) phase~\cite{Derrida80}.  Such transition was also corroborated by
$1/d$-expansions~\cite{Sourlas81}, as well as within mean-field and RG
for $m=2$ case~\cite{Continentino93}. However local mean-field
theory~\cite{Fischer87} predicts the breakdown of long-range FM order.

Following the arguments of Imry and Ma~\cite{Imry75} formulated for
the random-field model it was shown that the $d=3$ random anisotropy
magnet should break into magnetic domains of size
$L\sim({J}/D)^2$~\cite{Alben78} for weak anisotropy and thus no
ferromagnetism was expected. We will discuss this point in more
details in Section~\ref{subsect:IM}.

Account of fluctuations within the field theoretical RG
approach~\cite{amit} leads to an absence of phase transition.  In the
pioneer RG calculations performed for RAM with {\em isotropic
  distribution} of $\hat{\mathitbf{x}}_{\mathitbf r}$~\cite{Aharony75}
no stable accessible fixed point of the RG transformation was found in
the first order of the $\varepsilon=4-d$-expansion.  Moreover, the
effective Hamiltonian for such distribution at large $D$ was shown to
reduce to one that is similar to the effective Hamiltonian of the
random-bond Ising spin glass~\cite{Chen77}, demonstrating a
possibility of a SG phase in this case.

 Several arguments were used in order to demonstrate an absence of the
 FM order for space dimensions $d<4$ in RAM
 ~\cite{Pelcovits78}. Although among these arguments the one for the
 limit $m\to\infty$~\cite{Pelcovits78} appeared to be
 erroneous~\cite{Pelcovits82}, the lack of ferromagnetism for the RAM
 with {\it isotropic distribution} of anisotropy axes for $d<4$ was
 further supported by a Mermin--Wagner type proof~\cite{Pelcovits79}
 using the replica trick~\cite{Emery75} for the $m=2$ case. The
 perturbative Migdal-Kadanoff RG
 studies~\cite{Pelcovits78,Pelcovits82} also suggested dimensional
 reduction for RAM: critical behavior of this random system at $d>4$
 is the same as for the corresponding pure system with dimension
 $d-2$.

The one-loop  result~\cite{Aharony75} for the absence of the second-order
phase transition into ferromagnetic state was corroborated by
two-~\cite{Dudka01a,Dudka01c,Dudka02,Dudka05} and
five-loop~\cite{new_Ital} calculations within the field-theoretical RG
refined by resummation techniques.

The infinitely strong anisotropy limit of the RAM (which makes spins to be 
frozen in directions of local anisotropy axes and the Hamiltonian to be similar 
to that of the Ising random-bond spin-glass model) was
investigated with the help of high-temperature expansions. Results of a
Pad\'e-analysis~\cite{Shender80} indicated typical spin-glass behavior for
space dimension $d=3$, while in Ref.~\cite{Fisch90a} after obtaining non-power-law divergence of the susceptibility of three-dimensional RAM for $m=2$ and
no divergence in the case $m=3$ it was concluded that the lower critical
dimension for the RAM with $m=2$ is $d_{L}=3$.

The  series analysis of Ref.~\cite{Harris87} of the RAM in the infinitely strong anisotropy limit 
 on Cayley-trees predicted FM order, occurring for
the number of nearest neighbors $\tilde{z}> m$ and a SG order for
$\tilde{z}< m$. Results obtained in the same study~\cite{Harris87} by
Migdal-Kadanoff position space RG corroborate this outcome: $d=3$
FM order for small $m$, while for large $m$ the SG phase is
obtained in the same universality class as that of the Ising spin glass with randomly
distributed couplings.

Investigations of the RAM in the spherical model limit $m\to\infty$
were concentrated on the question about the possibility of a SG
phase. This limit was studied by
$1/m$-expansions~\cite{Goldschmidt83,Khurana84,Goldschmidt84,Jagannathan86}
first.  Within the replica method, a SG phase was found below $d=4$
for arbitrary $D$~\cite{Goldschmidt83,Boyanovsky83}. Later, the
spin-glass solution was shown to be unstable~\cite{Khurana84}. A
stable non-replica-symmetric solution was obtained for SG phase for
$d<4$~\cite{Goldschmidt84}. However, the study of dynamics of SG order
parameter avoiding the replica method for the RAM shows an instability of
the SG phase~\cite{Jagannathan86}. These results found their
confirmation in the mean-field treatment of the $m\to\infty$
limit~\cite{Fisher91}, where the SG phase appeared only as a feature
of this limit and no SG phase was obtained for finite $m$.  In the
spherical limit a FM order was obtained for $d>2$ and for $D$ less
than some critical value $D_c$, while for $D$ larger than $D_c$ a SG phase
was obtained for arbitrary $d$ ($D_c=0$ for $d\le
2$)~\cite{Khorunzhy91}.

An equation of state of the RAM showing a zero magnetization and an
infinite magnetic susceptibility in the low-temperature phase for any
$m$ was obtained perturbatively~\cite{Aharony80}. Two-spin
correlations in this phase possess a power law decay.  As mentioned
above, such phase {appears to be} QLRO. The estimate of the
susceptibility of the low-temperature phase was corrected using
scaling arguments~\cite{Aharony83} and it was found to be
$\chi\sim(D/{J})^{-4}$ at $d=3$. A similar dependence of the
susceptibility was obtained by other
approaches~\cite{Goldschmidt83,ChudnovskySaslowSerota}. The power law
decay of spin correlations in the low-temperature phase was obtained
in particular for harmonic system with random fields~\cite{Villian84},
which is an approximation of model with $m=2$ if vortices are
neglected. But the last result is in disagreement with
calculations~\cite{Dotsenko83} for model with random $p$-fold fields
(for two-component fields and $p=2$ it corresponds to the RAM with the
two-component order parameter) where the SG phase was obtained.

The QLRO ~\cite{Feldman99} was found for the RAM with $m=3$ by the
functional RG approach in the first order of $\varepsilon=4-d$
expansion. The functional RG study of higher rank anisotropies argued
that dimensional reduction breaks down~\cite{Fisher85}.  It was
corroborated within two-loop functional
RG~\cite{TissierTarjus,Doussal06}. These studies also showed that QLRO
exists for RAM at $d^*_{LC}<d<4$ and $m<9,4412$.  Estimates based on
expansions in $\varepsilon$ and $m-m^*$ predict $d^*_{LC}>3$, however
results obtained for small $\varepsilon$ and $m-m^*$ should be
extrapolated for large deviations with caution, as pointed out in
Ref.~\cite{Doussal06}. Conditions for holding dimensional reduction
were studied by $1/m$ expansions within the functional
RG~\cite{Sakamoto}.  In Ref.~\cite{Ideura} it was stated that all
previous studies using $1/m$ expansions are not completely correct,
since they do not take the large $m$ limit of the easy axes into
account. However only the case of $D<0$ was of main interest in that
study. Recent research of the large-$m$ limit reported glassy
character of the zero-temperature state~\cite{Mouhanna16}.

The phenomenological theory~\cite{ChudnovskySaslowSerota} based on
the continuous-field version of the Hamiltonian (Eq. (\ref{origham})) and
assuming correlations between randomly oriented an\-iso\-tro\-py
axes turned out  to be a more appropriate approach for the
interpretation of the field dependence of the experimentally
observed magnetization in the ordered phase. In this approach  the
spin correlation function in different regimes
 of applied fields was analyzed~\cite{ChudnovskySaslowSerota}. In particular, the
correlation length for small and zero fields was found to have the
form $\xi\sim R_a(\frac{{J}}{R_a D})^2$, where $R_a$ is the
correlation range for random axes. Such a phase was called a
correlated SG phase.

Once we have discussed the isotropic disorder distribution, we will discuss 
the {\em anisotropic} case. 

It was firstly investigated in a RG study~\cite{Aharony75} with a
distribution of anisotropic axes, restricting directions of the axes
along the hypercube edges ({\em cubic distribution}). No accessible
stable fixed point corresponding to a second-order phase transition
point was found. Despite this, the possibility of a second-order phase
transition into a FM phase with critical exponents of the
diluted quenched Ising model for the RAM with a cubic distribution was
pointed out in Ref.~\cite{Mukamel82}, where a more general case was
considered.  Such peculiarities were observed for a more general
model~\cite{Korzhenevskii88} including the RAM with {\em cubic
  distribution} of random axes as a particular case.  Subsequently,
this result was corroborated within a two-loop RG calculations with
resummation~\cite{Dudka01c,Dudka02,Dudka01b,Dudka05} done directly for
the RAM with {\em cubic distribution} showing that the critical
behavior belongs to the universality class of the site-diluted Ising
model. This result was further confirmed on the basis of a five-loop
massive RG calculations~\cite{new_Ital}. While study of RAM with
generic distribution~\cite{Mukamel82,Korzhenevskii88}, including {\em
  isotropic} and {\em cubic} distributions as particular cases,
{within the RG approach} followed by resummation has predicted
continuous phase transition of a new universality class~\cite{Dubs17},
later it was demonstrated that this conclusion was based on 
erroneous calculations~\cite{Shapoval}.

The RAM in the infinite anisotropy limit with
mixed isotropic and cubic distributions was also investigated by
mean-field theories~\cite{Fischer85,Fischer86,Dominguez93}.  It was
found that the presence of a random cubic anisotropy stabilizes the
FM phase~\cite{Fischer85}.  Study in the limit $m\to
\infty$ with finite $\alpha\equiv m/N$ , where $N$ is number of spins
(so called $\alpha$-limit)~\cite{Dominguez93} gives phase diagrams
with FM and  SG phases as well as with mixed phase
where both FM and SG order parameters are nonzero.

Summing up, from the analysis of theoretical results one may conclude
that while for an isotropic distribution the absence of FM ordering is
expected, such ordering is possible for anisotropic distribution of
the local anisotropy axis.

\subsection{Numerical results}

Most of the performed numerical simulations of the $d=3$ RAM study the
cases $D/{ J }>1$ or more often the infinitely strong anisotropy ($D/{
  J }\to\infty$) limit.
  
The earliest investigations report inconclusive results: both
stability~\cite{Chi77,Harris78} as well as
instability~\cite{Chi79,Harris80} of the FM order with respect to the
SG phase have been reported.  However, data of later investigations
indicated the absence of ferromagnetism.

The restriction to the infinitely strong anisotropy limit led to a
lack of long-range order in the ground state for
$m=3$~\cite{Jayaprakash80}. In this case, the critical exponents at
the transition to a low-temperature phase were reported to be similar
to those of the three-dimensional short-range Ising spin
glass~\cite{Chakrabati87}. Results of Monte Carlo
simulations~\cite{Fisch89} confirmed an absence of long-range
order. Finally, the  most recent study of the infinitely strong anisotropy
limit of RAM with $m=3$ convincingly shows that the model in this case
belongs to the class of short-range Ising spin-glass with bond
disorder~\cite{ParisienToldin06,Liers07}.

On the other hand, results of Monte Carlo study of the infinitely strong anisotropy
limit of RAM with $m=2$ show a low-temperature phase
with extremely large susceptibility, power law decay of the correlations
and vanishing magnetization~\cite{Fisch89}, which  is consistent with
the theoretically  predicted QLRO for arbitrary $m$ and
weak anisotropy~\cite{Aharony80}. That is confirmed by the result of
Ref.~\cite{Fisch91}, where a sharp phase transition into a 
low-temperature phase with power-law decay of the correlation
function and no true magnetization were found. This case was also
studied in the infinite anisotropy
limit~\cite{Fisch90} of a model with random twofoldfold fields, where weak FM order was found with
power-law correlations.

Numerical simulations of RAM with finite ratio of $D/J$ reported the
phase transition into SG phase for the model with random twofold fields
at ${J}=D$~\cite{Reed91}. Another Monte Carlo study of the RAM
(\ref{origham}) with $m=2$ and $J=D$ resulted in critical exponents
with values similar to the XY-ferromagnetic transition, except that
the heat capacity critical exponent was found to be
positive~\cite{Rossler}.

Performing Monte Carlo calculations for the RAM Hamiltonian
(\ref{origham}) with $m=3$ and several values of $D/J$ as well as at
$D/J \to \infty$ the phase diagram in the plane $(D/J,T/J)$ was
found~\cite{Itakura03}.  There, the regions of existence of magnetic
and SG order were indicated, with general conclusion that a random
anisotropy Heisenberg model for small $D/J$ has a QLRO low-temperature
phase characterized by frozen power law spin correlations.

Results of study of the model with random twofold fields at several
values of $D/J$ suggest that system is ferromagnetic in low-temperature phase at finite values of $D/J$~\cite{Fisch2009}.

Results reviewed above concern cases of continuous symmetry of the
order parameter. Below we consider cases where orientations of spins
as well as of the local anisotropy axes are limited only to several
directions in $m$-dimensional space.  In particular, studying RAM with
$m=2$, where the spins and anisotropic axes are oriented along the
edges of a cube, the conventional XY second-order phase transition to
the FM phase was found for weak random anisotropy~\cite{Fisch93},
whereas a first-order transition to a domain type FM phase was found
for strong random anisotropy. For $m=3$ both transitions were found to
be of the first order~\cite{Fisch93}.

The possibility of the existence of a QLRO phase was
also obtained for $m=3$ in the case of weak anisotropy but assuming
$D/J=\infty$ for a part $q$ of sites and $D/J=0$ for the rest of
$1-q$ sites~\cite{Fisch98}.  There, the spins and anisotropic axes were
chosen from the 12 directions.  Results indicate, that in addition to
paramagnetic (PM) and the FM ordering a QLRO
phase appears as an intermediate phase for some values of $q$.

The RAM with two-component spins is adopted to describe six-state clock model,
where direction of spins belongs to the $Z_6$ group, while the local anisotropy
orientation is taken from the $Z_3$ group. The obtained low-temperature phase in
this model has two-spin correlations decaying according to a power law, but no
long-range magnetic order~\cite{Fisch95}.
 
To summarize, the reviewed numerical studies agree, in general, with
the theoretical predictions about the absence of long-range order for
the $d=3$ RAM with isotropic random axes distribution. However, they
show the possibility of a FM order for the cases where orientations of
local anisotropy axes are limited only to several directions.

An exception is given by the study of RAM with Heisenberg spins
($m=3$)~\cite{Nguyen09}. Studying the case $D/J{=}4$ for {\em
  isotropic} distribution and {\em cubic} distributions~ surprisingly
second-order phase transition to ferromagnetically ordered state
(mistakenly identified as QLRO) was found with the same correlation
length critical exponent.  To check this outcome we perform similar
computations for larger lattices and for two values of $D/J$. But
before describing our model let us go back to the cornerstone of
the argumentation of an absence of long-range order for $d=3$ RAM, the
Imry-Ma arguments.

\subsection{Imry-Ma arguments}
\label{subsect:IM}

Imry and Ma introduced domain arguments in order to understand the low-temperature phase of random field magnets~\cite{Imry75}. Later,
similar arguments were directly applied to the random anisotropy
case~\cite{rev,Alben78,Feldman99}.

The existence of random directions in the RAM system can lead to its 
splitting into domains inside which spins are directed almost along
one direction. If the typical size of the domains is $L$, than energy
gain for system is $\sim D L^{d/2}$. {Whereas the loss of the surface
  energy per domain for the continuous symmetry order parameter can be
  estimated as $\sim J L^{d-2}$.}  Minimizing total energy from these
two contributions with respect to $L$ one gets
$L\sim\left(J/D\right)^{2/4-d}$. That means that even for very small
anisotropy strength RAM always should split into domain for $d<4$.
An earlier similar result was obtained by Larkin~\cite{Larkin} in the
context of a vortex lattice in a superconductor. Therefore one can find
different names of the typical size of domains: Larkin length, Imry-Ma
length or Imry-Ma-Larkin length.

Validity of these arguments was questioned. Berzin, Morosov and
Sigov~\cite{Berzin16} pointed that random-field arguments can not be
directly applied to the random-anisotropy case since the orientations
of $\hat{\mathitbf{x}}_{\mathitbf r}$ in some direction of the order
parameter space and opposite to it are equivalent in RAM. Moreover
considering continuous field variant of the RAM they have shown that
the long-range order is possible when the distribution of
$\hat{\mathitbf{x}}_{\mathitbf r}$ deviates from the isotropic one as
well as the local anisotropy axis is present only on part of the 
sites~\cite{Berzin}.

Fisch~\cite{Fisch2009} pointed out that with the presence of random vectors
$\hat{\mathitbf{x}}_{\mathitbf r}$ the random anisotropy system is not
longer translationally invariant. Therefore the consideration that the
twist energy at the boundary scales in the same way as in nonrandom
magnets, i.e., $\sim L^{d-2}$, is not correct.

Nevertheless, as it can be seen below our results are in agreement
with Imry-Ma arguments about the existence of the Larkin-Imry-Ma length and
the absence of ferromagnetism for RAM with isotropic distribution.

\section{The Model and Simulations} \label{III}

We have studied the Hamiltonian given by Eq. (\ref{origham}) with
$m=3$ on a three-dimensional lattice of linear size $L$ and volume
$V=L^3$ with periodic boundary conditions.

\subsection{Random axis distributions}

Two kinds of disorder have been simulated. In the first case, called
hereafter ``isotropic disorder'', the random anisotropy vectors
$\hat{\mathitbf x}_{\mathitbf r}$ have an uniform probability
distribution on the sphere of unit radius
\begin{equation}
   p(\hat{\mathitbf{x}}) = \frac{1}{4 \pi}\, .
\end{equation}
 The results obtained for such disorder
will be called in the figures and tables IRAM.

In the second case, called cubic or ``anisotropic disorder'' and called
hereafter ARAM, the vectors $\hat{\mathitbf x}_{\mathitbf r}$
point along the six semi-axes of the cubic lattice with the same
probability $1/6$:
\begin{equation}
       p(\hat{\mathitbf{x}})=\frac{1}{6} \sum_{i=1}^3 \left( \delta
    (\hat{\mathitbf{x}},\hat{\mathitbf{k}}_i)
    + \delta
    (\hat{\mathitbf{x}},-\hat{\mathitbf{k}}_i)\right)\,,
\end{equation}    
where the $\hat{\mathitbf{k}}_i$ ($i=1,2,3$) are the three unit
vectors pointing to the three principal directions of a cubic lattice
and $\delta(\hat{\mathitbf{x}},\hat{\mathitbf{k}}_i)$ are Kronecker
deltas.  In our numerical calculations we fix $J=1$, therefore
anisotropy strength determines the ratio $D/J$. We measure temperature
$T$ in units of Boltzmann constant $k_\mathrm{B}$ and work mainly with
the inverse temperature $\beta=1/T$.

We have run Monte Carlo numerical simulations using the Metropolis
(with 10 hits) and parallel tempering algorithms, see
Refs.~\cite{PT1996,Marinari1998}. See Appendix~\ref{num} for more
details on the parameters used in our runs and on the thermalization
tests. In addition to generate the isotropic distribution we have used
the standard rejection method~\cite{harman:10}.

\subsection{Observables}

In this section we will describe the observables we have simulated.
The magnetization is computed as
\begin{equation}
M=\overline{\left\langle \sqrt{\boldsymbol{\mathcal M}^2}\right\rangle} \, ,
\label{orderparameter}
\end{equation}
with
\begin{equation}
\boldsymbol{\mathcal M} =\frac{1}{V}\sum_\mathitbf{r}
\mathitbf{S}_\mathitbf{r} \,,
\label{mag}
\end{equation}
where, as usual, $\langle(\cdot \cdot \cdot)\rangle$ is the thermal
average and $\overline{(\cdot \cdot \cdot)}$ denotes the average over
the disorder.

Calculating powers of $\boldsymbol{\mathcal M}$ we can get the associated susceptibility 
\begin{equation}
\chi=V\overline{\left\langle \boldsymbol{\mathcal M}^2 \right\rangle}\, ,
\label{susceptibility}
\end{equation}
and the Binder cumulant
\begin{equation}
U_4=1-\frac{1}{3}\frac{\overline{\langle (\boldsymbol{\mathcal M}^2)^2\rangle}}
           {\overline{\langle \boldsymbol{\mathcal M}^2 \rangle}^2} \, .
\label{binder}
\end{equation}
With this definition the Binder cumulant will be asymptotically zero
in a paramagnetic phase (Gaussian probability distribution of the
magnetization with zero mean), and equal to $2/3$ in a ferromagnetic
phase.

A definition of the correlation length on a finite lattice
is~\cite{XIL,amit}:
\begin{equation}
\xi=\left(\frac{\chi/F-1}{4\sin^2(\pi/L)}\right)^\frac{1}{2}\,,
\label{XI}
\end{equation}
where $F$ is defined as the Fourier transform of the
magnetization via
\begin{equation}
  \boldsymbol{{\cal F}}(\mathitbf{k})=
  \frac{1}{V}\sum_{\mathitbf{r}} e^{\mathrm i
\mathitbf{k}\cdot\mathitbf{r}} \mathitbf{S}_\mathitbf{r}
\end{equation}
with
\begin{equation}
F=\frac{V}{3}\overline{\left\langle |\boldsymbol{\mathcal
F}(2\pi/L,0,0)|^2+\mathrm{two~permutations}\right\rangle}\, .
\end{equation}

The interpretation of $\xi$ as defined in Eq. (\ref{XI}) is the
following (see for more details section 2.3.1 of Ref. \cite{amit}): i)
In the paramagnetic region $\xi\to \xi_2^\infty$ for large $L$, where
$\xi_2^\infty$ is the second-moment correlation length computed in an
infinite volume which is proportional to the exponential correlation
length, ii) in the critical region $\xi\propto L$ and iii) in the
ferromagnetic phase $\xi\sim L^{d/2}$ since $F={\cal O}(1)$ and
$\chi\sim {\cal O} (L^d)$.

Therefore $\xi$ will converge to the true correlation length only for
$T\ge T_c$. In the low-temperature phase it will diverge and it is no
longer the correlation length of the ferromagnetic phase. However,
this fact is used in numerical studies to characterize the phase
transition~\cite{amit}. The cumulant $R_\xi=\xi/L$ will tend to zero
(decreasing with $L$) in the paramagnetic region, and will diverge
(growing with $L$) in the ferromagnetic region. Thus, the curves of
$R_\xi$ for different lattice sizes will cross near the critical
point.

For a finite system of size $L$, the Binder cumulant obeys:
\begin{equation}
U_4(\beta,L) = f((\beta-\beta_c)L^{1/\nu} )+O(L^{-\omega}),
\end{equation}
where $\nu$ is the correlation length critical exponent, $\omega$ is
the leading correction to scaling and $f(\cdot)$ is a scaling
function.  Neglecting scaling corrections, the curves of the Binder
cumulant as a function of the temperature for two different lattice
sizes intersect in one point, indicating the approximate location of
critical temperature $\beta_c$. The same behavior is expected for the
dimensionless cumulant defined as a ratio of the correlation length to
system size, $R_\xi=\xi/L$.

We have also computed the derivatives of the Binder cumulant
($\partial_\beta U_4$) and the correlation length ($\partial_\beta
\xi$) by computing connected average values of different moments of
the magnetization (and $\cal F$ in the case of $\xi$).  The derivative
for quantity $ {\cal O}$ is calculated with the help of the total
energy ${\cal E}$ defined by Eq.(\ref{origham}):
\begin{equation}
\partial_\beta \overline{\langle {\cal O}\rangle}=
 \overline{\partial_\beta\langle {\cal O}\rangle}=
\overline{\left\langle_{\vphantom{|}} {\cal{OE}} - \langle{\cal O}
 \rangle \langle {\cal E} \rangle\right\rangle}.
\end{equation}

We have also corrected the bias of having a relatively small
number of measures per sample: to do that we have applied a third
order extrapolation as described in Ref.~\cite{Extra}.

To measure the lack of self-averaging we have computed the $g_2$ cumulant, defined as
\begin{equation}
g_2=\frac{ \overline{\langle  \boldsymbol{\mathcal M}^2  \rangle^2}-\overline{\langle
    \boldsymbol{\mathcal M}^2  \rangle}^2}{\overline{\langle \boldsymbol{\mathcal M}^2  \rangle}^2}\,.
\end{equation}
If $g_2$ goes to zero when system size $L$ increases then
susceptibility is a self averaging quantity. Otherwise the
susceptibility does not self average (for details, see e.g.,
Ref.~\cite{Wiseman95}).

To analyze the spin-glass behavior in the isotropic disorder we have
computed the overlap (see for example Ref. \cite{Campos97})
\begin{equation}\label{overlap}
    q_{\boldsymbol{r}}^{\alpha \beta}=S^{(1),\alpha}_{\boldsymbol{r}} S^{(2),\beta}_{\boldsymbol{r}} \,,
\end{equation}
where ${S}^{(1), \alpha}_{\boldsymbol{r}}$ and ${S}^{(2),
  \beta}_{\boldsymbol{r}}$ ($\alpha,\beta=1,2,3$) are the components
of two spin replicas $\boldsymbol{S}^{(1)}_{\boldsymbol{r}}$ and
$\boldsymbol{S}^{(2)}_{\boldsymbol{r}}$ which evolve with the same
disorder\footnote{We have checked that the overlap computed as the
scalar product of the spins belonging to different replicas, shows the
same properties as the one defined in the text.}.

The Fourier transform of the tensor overlap reads
\begin{equation}
  \boldsymbol{{\cal F}}_q^{\alpha\beta}(\mathitbf{k})=
  \frac{1}{V}\sum_{\mathitbf{r}} e^{\mathrm i
\mathitbf{k}\cdot\mathitbf{r}} q_{\mathitbf{r}}^{\alpha \beta}\,,
\end{equation}
In turn, the spin-glass susceptibility is
\begin{equation}
\chi_q   = V  \sum_{\alpha\beta}\overline{\left\langle |\boldsymbol{\mathcal
F}_q^{\alpha,\beta}(\boldsymbol{0})|^2\right\rangle}\,,
\end{equation}
and the correlation length in the spin-glass channel is
\begin{equation}
\xi_q=\left(\frac{\chi_q/F_q-1}{4\sin^2(\pi/L)}\right)^\frac{1}{2}\, ,
\label{XIq}
\end{equation}
where
\begin{equation}
F_q=\frac{V}{3}\sum_{\alpha, \beta}\overline{\left\langle |\boldsymbol{\mathcal
F}_q^{\alpha\beta}(2\pi/L,0,0)|^2+\mathrm{two~permutations}\right\rangle}\,.
\end{equation}
As for the magnetization, we can define the following dimensionless observable in the spin-glass channel: $R_{\xi_q}=\xi_q/L$.

The behavior of $R_{\xi_q}$ is analogous to that of $R_\xi$ (see above
in the text): it will go to zero in the paramagnetic phase and it will
diverge in the spin-glass one as $L$ increases. The crossing of the
different lattice size curves will mark the critical point.

\section{Results and Analysis} \label{IV}

\begin{figure}[hbt!]
\centering
\includegraphics[width=\columnwidth, angle=0]{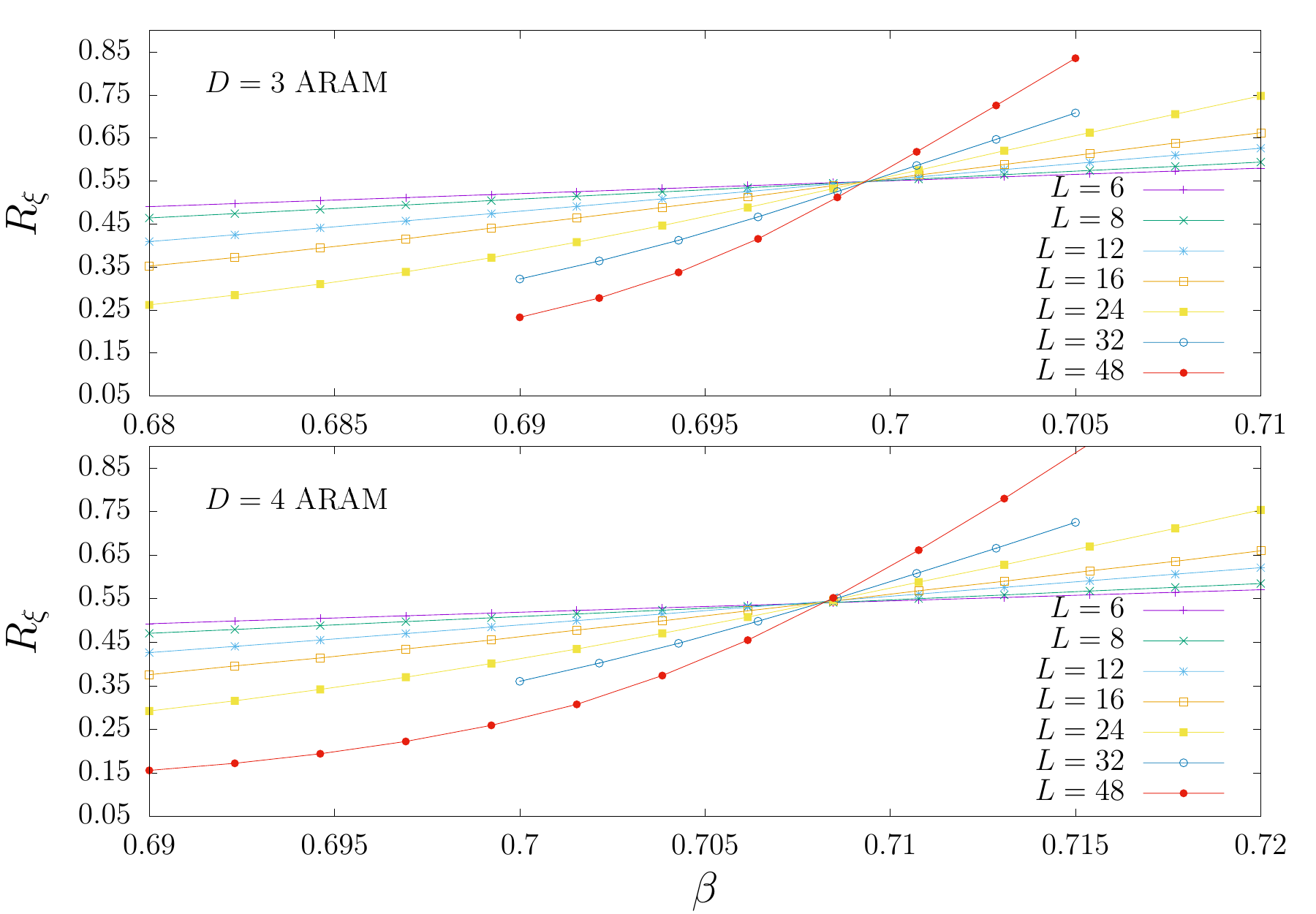}
\caption{(color online) $R_\xi$ cumulant versus inverse temperature, $\beta$, for the cubic disorder for several lattice sizes.}
\label{fig:XILARAM}
\end{figure}

\begin{figure}[ht]
\centering
\includegraphics[width=\columnwidth, angle=0]{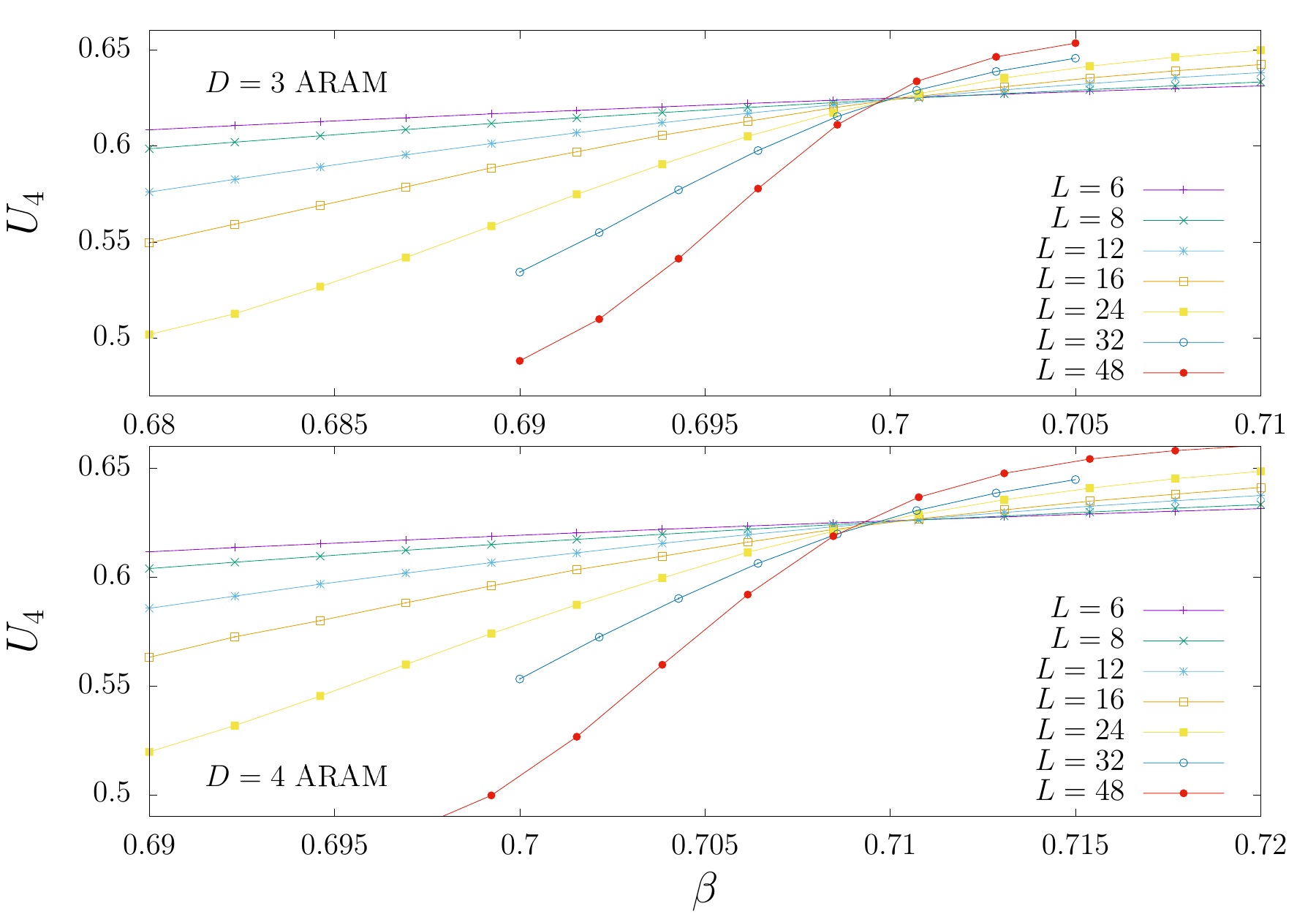}
\caption{(color online) Binder cumulant versus inverse temperature for the cubic disorder for several lattice sizes.}
\label{fig:G4ARAM}
\end{figure}

We have computed the crossing points of the $R_\xi(\beta,L)$ or Binder
curves of $L$ and $2 L$ lattices:
$R_\xi(\beta_\text{cross}(L,2L),L)=R_\xi(\beta_\text{cross}(L,2L),2L)$
or $U_4(\beta_\text{cross}(L,2L),L)=U_4(\beta_\text{cross}(L,2L),2L)$.
We have used a fifth-order polynomial-based analysis to compute the
crossing temperatures.

Once we have found the phase transitions in the four models, we have
analyzed quantitatively the behavior of the different observables
using the quotient and the fixed coupling methods (see the Appendix
\ref{quotient} for a description of these methods).

Furthermore, we have computed the value of $R_\xi$ (or $R_{\xi_q}$ in
the case of a spin-glass phase transition) at $\beta_\text{cross}$,
the value of $\nu_\xi$ from the behavior of $\partial_\beta \xi$ (the
behavior of $\partial_\beta U_4$ will provide an additional estimate
of the $\nu$ exponent denoted as $\nu_{U_4}$). Finally, we have also
computed $\eta$ from the behavior of the susceptibility ($\chi$) and
also the ratios $Q_{U_4}$ and $Q_{g_2}$.

In the next two subsections we will discuss in detail our findings for the two disorder distributions.

\begin{figure}[ht]
\centering
\includegraphics[width=\columnwidth, angle=0]{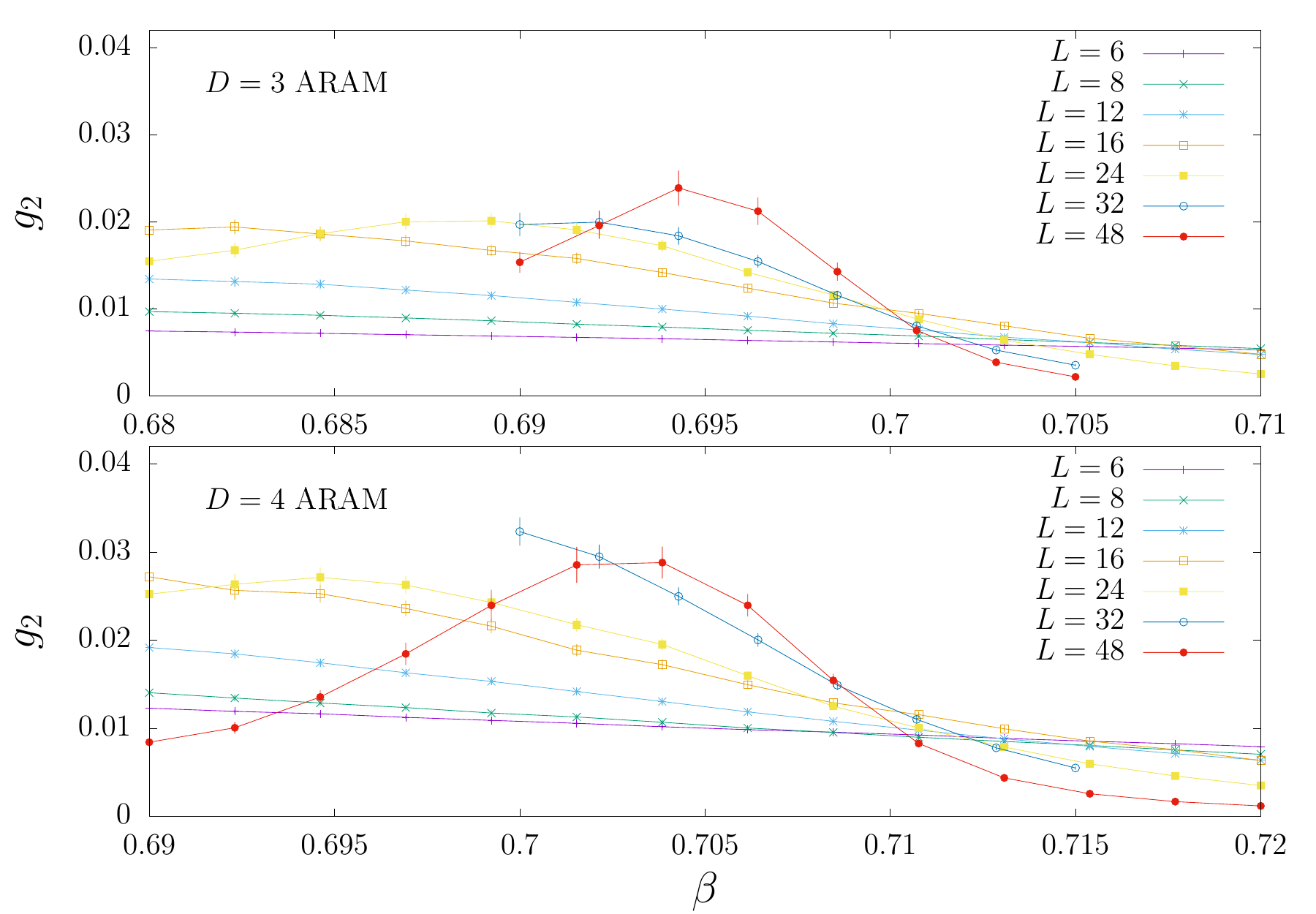}
\caption{(color online) $g_2$ cumulant as a function of the inverse
  temperature for the cubic disorder for several lattice sizes.}
\label{fig:G2ARAM}
\end{figure}

\begin{table*}[hbt!]
\caption{Quotient method results for  $D=3$ and $D=4$ and anisotropic 
  disorder  distribution (ARAM) from the crossing points of $R_\xi$ for lattice sizes $L_1$
  and $L_2$.}  \centering
\label{tab:resultsQA}
\begin{tabular}{|c|c|c|c|c|c|c|c|c|}
\hline
$D$ & $L_1/L_2$  & $\beta_\text{cross}$ & $R_\xi$ & $\nu_\xi$ & $\nu_{U_4}$ &
  $\eta$ & $Q_{U_4}$ & $Q_{g_2}$ \\
\hline\hline

3&  6/12&  0.6992(2)&  0.5485(6)&  0.756(1)&  0.747(2)& 0.019(3) &  0 9975(2)& 1.31(4)\\
3&  8/16 & 0.6993(2)&  0.5486(9)&  0.757(5)&   0.77(1)   & 0.026(5) &  0.9978(3)& 1.44(5)\\
3&  12/24& 0.69940(8) &  0.550(1)& 0.740(4)&   0.762(7)    & 0.031(5) &  0.9975(3)& 1.30(5)\\
3& 16/32& 0.69941(7) &  0.549(1)& 0.726(6)&   0.75(1)   & 0.030(6) &  0.9980(4)& 1.00(5)\\
3& 24/48& 0.69932(7) &  0.548(2)& 0.737(8)&  0.76(2)   & 0.03(1)&  0.9983(6)& 1.12(6)\\\hline\hline

4&6/12 & 0.7073(2) & 0.5383(7)  & 0.740(2)&  0.749(4) &   0.027(4) & 0.9954(2)& 1.17(3)\\
4&8/16 & 0.7079(2) & 0.540(1) &  0.751(7) & 0.78(1)&   0.035(6) & 0.9952(4)& 1.38(5)\\
4&12/24& 0.7083(1) &  0.545(1) & 0.734(5) & 0.76(1)  &  0.033(7) & 0.9962(5)&  1.17(5) \\
4&16/32& 0.7082(1) &  0.542(2) & 0.710(8) & 0.71(2)  & 0.036(9) & 0.9943(7)& 1.20(6)\\
4&24/48& 0.70830(7) &  0.545(2)  &  0.733(9) & 0.76(2)   &0.04(1) & 0.9947(9) &  1.26(7) \\\hline\hline
\end{tabular}
\end{table*}

\subsection{Analysis of the cubic disorder}
\label{IVA}

In Fig. \ref{fig:XILARAM} we show the behavior of $R_\xi$ as a
function of the inverse temperature $\beta$ for the anisotropic
disorder and $D=3$ and 4. We also show the behavior of the Binder
cumulant in Fig. \ref{fig:G4ARAM}: the signature of a phase transition
is really strong for both observables, Binder cumulant and correlation
length.  This is manifested by the crossing of the curves for
different lattice sizes near the critical point.

In addition, we have also plotted in Fig. \ref{fig:G2ARAM} the $g_2$
cumulant.  This cumulant shows clear crossing points (but much noisier
than $R_\xi$ and $U_4$), which do not extrapolate to zero (see below).
This is a strong signature that the model with cubic disorder does not
belong to the same universality class as a pure (nondisordered) model
($g_2=0$).

We start describing our findings obtained using the quotient method.

The Binder cumulant analysis presents strong scaling corrections,
therefore, we will present the results of the critical parameters
obtained with the analysis of the crossing temperatures of $R_\xi$.

The data presented in Table \ref{tab:resultsQA} show almost negligible
(in our numerical precision) corrections to scaling. Hence, we take
the results for the largest pair (24 and 48) as our final estimate
using the quotient method. We present the critical temperature,
$R_\xi$ cumulant at the critical point and correlation length and pair
correlation function critical exponents, $\nu$ and $\eta$, for $D=3$
and $D=4$ in Table \ref{tab:extrapolationQA}.  The nonmonotonic
behavior of the cumulants $Q_{g_2}$ and $Q_{U_4}$ (see Table
\ref{tab:resultsQA}), for both values of $D$, precludes us from computing
the $\omega$ exponent.
 
 We have also analyzed our numerical data using the fixed coupling
 method. We present our estimates in Table
 \ref{tab:extrapolationFCAA}. In this analysis we have used only the
 leading term: as in the quotient method we have been unable to
 characterize corrections to scaling (see Appendix \ref{quotient}).
 We would like to stress that the quotient and fixed coupling
 extrapolations are fully statistical compatible for each value of
 $D$. In addition, as a check of universality, it is important to
 state the statistical compatibility of all the four sets of
 extrapolated values (two different methods and two different values
 of $D$).  The behavior of $Q_{g_2}$ as a function of the lattice size
 (see Table \ref{tab:resultsQA}) is a strong hint for an asymptotic
 non zero value of $g_2$ at the critical point, hence, ruling out that
 the ARAM belongs to the (pure) Heisenberg universality class for
 $D=3$ and 4.

\begin{figure}[hbt!]
\centering
\includegraphics[width=\columnwidth, angle=0]{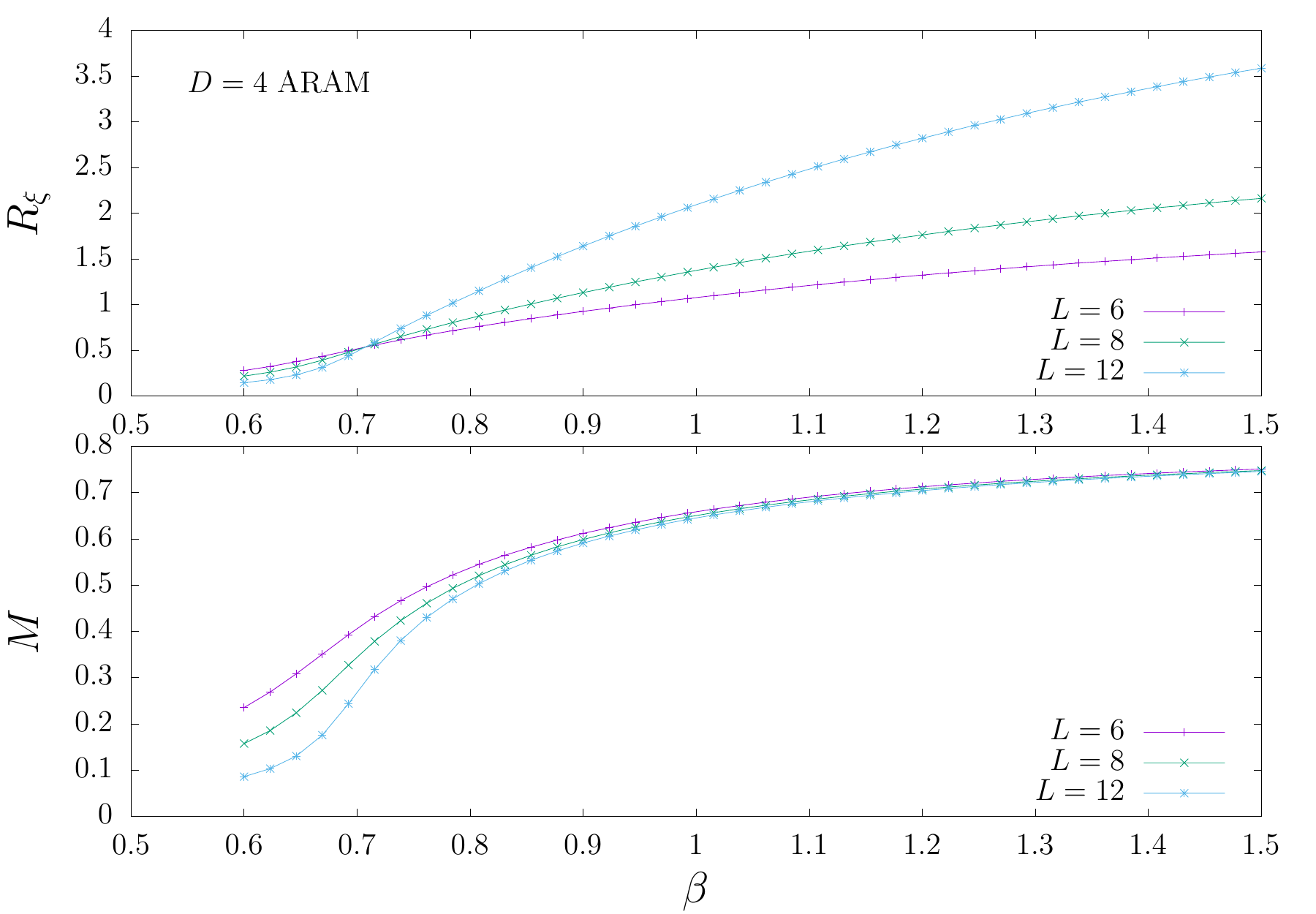}
\caption{(color online) $R_\xi$ cumulant (top) and $M$ (bottom) for
  $D=4$ and cubic disorder versus the inverse temperature for the
  smallest simulated lattices covering a wide range of
  temperatures. There is no additional phase transitions in this wide
  range of temperatures in the magnetic channel. The magnetization $M$
  is already asymptotic for lower temperatures and small lattice sizes
  and is clearly non zero.}
\label{fig:XILARAMEXT}
\end{figure}

Finally we have explored the behavior of the system at very low
temperatures in order to check the system remains in the FM phase, see
Fig. \ref{fig:XILARAMEXT}.  Notice that in the FM phase $R_\xi$
diverges with the lattice size, as is the case. Moreover, the
magnetization is asymptotic (clearly non zero), even for small lattice
sizes, in the low-temperature region.

\begin{table*}[ht!]
  \caption{ Inverse critical temperature, exponents and $R_\xi$ for
    the cubic disorder (ARAM), $D=3$ and 4, obtained with the quotient
    method analysis of the $R_\xi$-channel. As explained in the text,
    the Hamiltonian for $D=3$ and 4 shows small corrections to
    scaling, and we have quoted the results obtained from the biggest
    lattices (24 and 48).}  \centering
\label{tab:extrapolationQA}
\begin{tabular}{|c|c|c|c|c|c|}
\hline
$D$  & $\beta_\text{c}$ & $R_\xi$ &  $\nu_\xi$ & $\nu_{U_4}$ & $\eta$ \\
\hline
3 & 0.69932(7)  & 0.548(2)    & 0.737(7) & 0.76(2) &0.03(1)  \\
4 & 0.70830(7)  & 0.545(2)    & 0.733(9) & 0.76(2) &0.04(1)  \\ \hline

\end{tabular}
\end{table*}

\begin{table*}[ht!]
\caption{Fixed coupling method results
  (as described in the Appendix \ref{quotient}) for the cubic disorder (ARAM) and 
  $D=3$ and 4. We have used $R_\xi=0.544$ for both values of $D$.}  \centering
\label{tab:extrapolationFCAA}
\begin{tabular}{|c|c|c|c|c|c|}
\hline
$D$  & $R_\xi$ & $\beta_\text{c}$ &  $\nu_\xi$ &  $\nu_{U_4}$ &$\eta$ \\
\hline
3 & 0.544  & 0.69941(5)    & 0.733(3) &  0.76(5) & 0.028(1)\\
4 & 0.544  & 0.70827(5) & 0.736(3) &  0.72(1)    & 0.035(3) \\ \hline
\end{tabular}
\end{table*}

\subsection{Analysis of the isotropic disorder}
\label{IVB}
\subsubsection{Magnetization channel}

Let us start the study of the isotropic disorder simulating in the
region, $\beta\sim 0.7$, where a phase transition was found for this
disorder in the magnetic channel, for $D=4$, in Ref.~\cite{Nguyen09}.

\begin{figure}[hbt!]
\centering
\includegraphics[width=\columnwidth, angle=0]{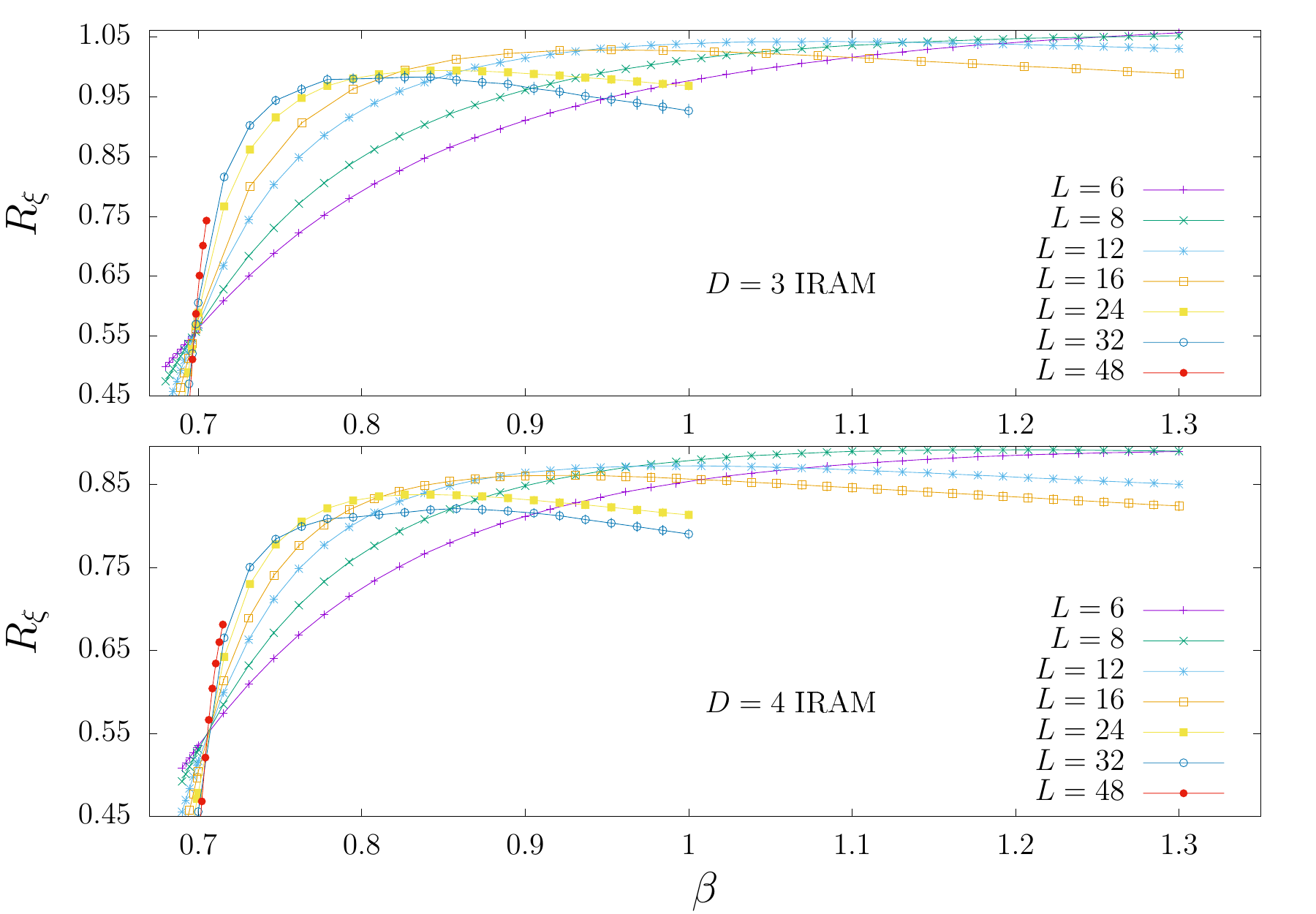}
\caption{(color online) $R_\xi$ cumulant as a function of the inverse temperature for the isotropic disorder for several lattice sizes at $D=3$ (top) and $D=4$ (bottom).}
\label{fig:XILIRAMEXTD34}
\end{figure}

The results are shown in Fig. \ref{fig:XILIRAMEXTD34}. As one can see
from this figure, coming from the PM region (from higher to lower
temperatures), all the $R_\xi$ cumulant curves cross in the
neighborhood of $\beta_c^1\sim 0.7$ for all the simulated lattice
sizes.  This is an evidence of a PM-FM continuous phase transition. A
similar behavior for the Binder cumulant was also found in
Ref.~\cite{Nguyen09}.

The crossing points of the $R_\xi$-curves are very stable as the
lattice size is growing, see Table \ref{tab:cross_merg}. This fact
allows us to determine the crossing temperatures, further denoted as
$\beta_c^1(L,2 L)$, and to proceed to calculate the critical exponents
using different observables. The results are described in Appendix
\ref{ISO}. However, these exponents are to be considered as
``effective'' ones for reasons we will explain below.
  
We have extended the numerical simulations, from the region $\beta\sim 0.7$ to
the very low-temperature one and, surprisingly, we have detected a
clear signature of a second phase transition (marked again by the
crossing points of the $R_\xi$ curves, see
Fig. \ref{fig:XILIRAMEXTD34})\footnote{We thank an anonymous referee
for pushing us to look at the low-temperature region.} between a FM
phase and one that we will denote as the  zero magnetization phase
(ZM), see below for the justification of this name.  Furthermore, we
denote the crossing points of this phase transition by
$\beta_c^2(L,2L)$ and we have computed them in Table
\ref{tab:cross_merg}.
 
A remarkable feature of the curves in Fig. \ref{fig:XILIRAMEXTD34} is
that whereas the crossing points $\beta_c^1(L,2 L)$ of the first phase
transition are stable with respect to changes in $L$, the crossing
points corresponding to the second phase transition, $\beta_c^2(L,2
L)$, show a huge drift towards the high temperature region (i.e., towards the
first phase transition). Moreover, $\beta_c^2(L,2 L)$ converges to
$\beta_c^1(L,2 L)$ as the lattice size $L$ grows, as one can see in
Fig.  \ref{fig:XILIRAMEXTD34}.

  In Table \ref{tab:cross_merg} we report the difference between the
  crossing points computed as a function of the lattice size for the
  observed PM-FM and FM-PM transitions. For both values of $D$ the
  differences scale to zero following a power law: $\beta_c^2(L,2
  L)-\beta_c^1(L,2 L) \propto L^{-a}$, finding very good fits (for
  $L>6$) for both values of $D$, with the exponent $a\simeq 1.5$
  ($D=4$) and $a\simeq 1.7$ ($D=3$).
  
\begin{figure}[hbt!]
\centering
\includegraphics[width=\columnwidth, angle=0]{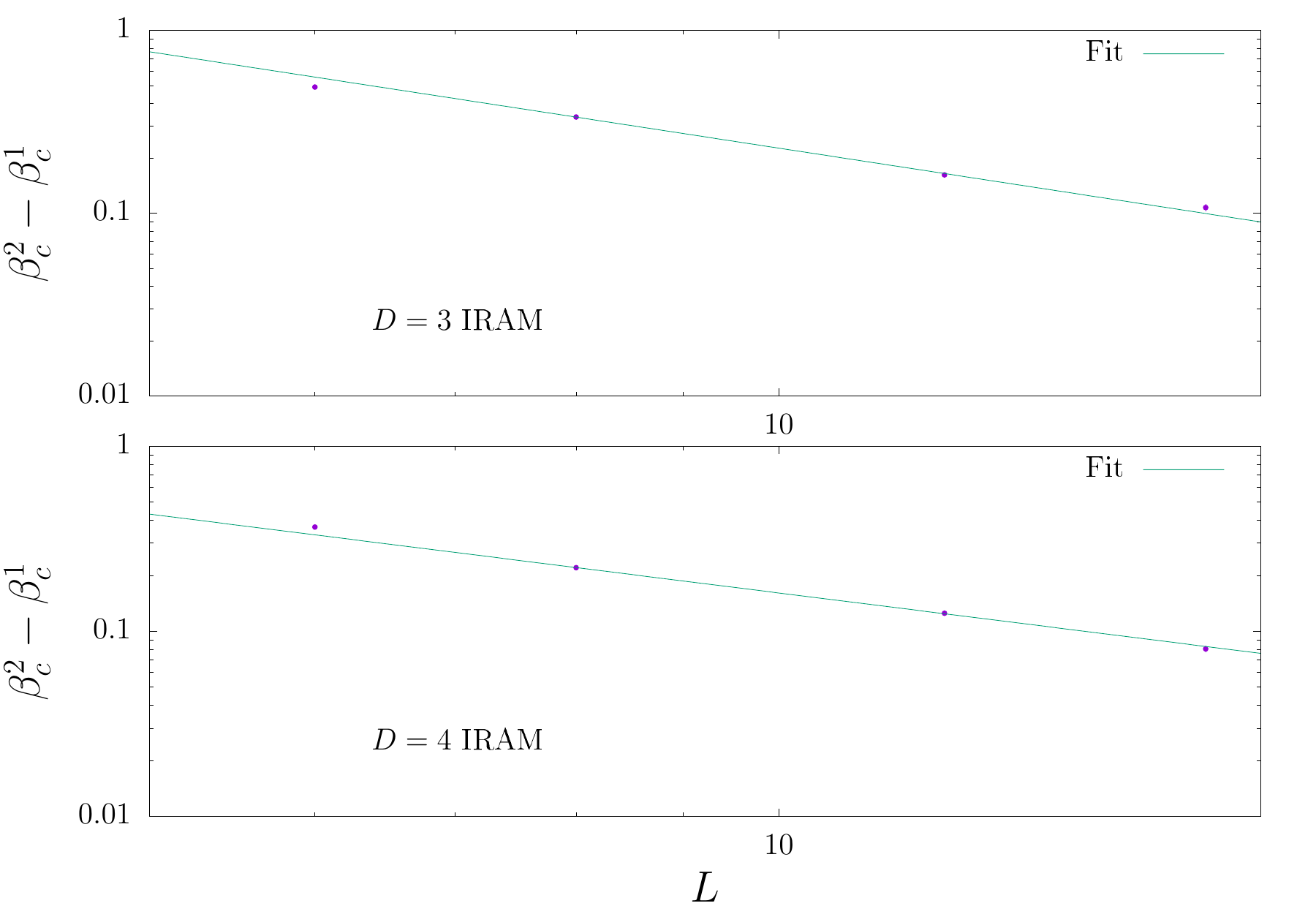}
\caption{(color online) Difference of the crossing points of the
    first (PM-FM), $\beta_c^1(L, 2L)$, and the second (FM-ZM), $\beta^2_c(L, 2L)$, 
    transition as a function of $L$ for $D=3$ (top) and $D=4$
    (bottom). The continuous lines are the fits to a pure power law,
    $\beta_c^2-\beta_c^1 \propto L^{-a}$.}
\label{fig:DIFF}
\end{figure}

  In Fig. \ref{fig:DIFF}, the difference between the crossing points of the two
  transitions is plotted as a function of the lattice size, where we have also
  plotted the pure power law.

\begin{figure}[hbt!]
\centering
\includegraphics[width=\columnwidth, angle=0]{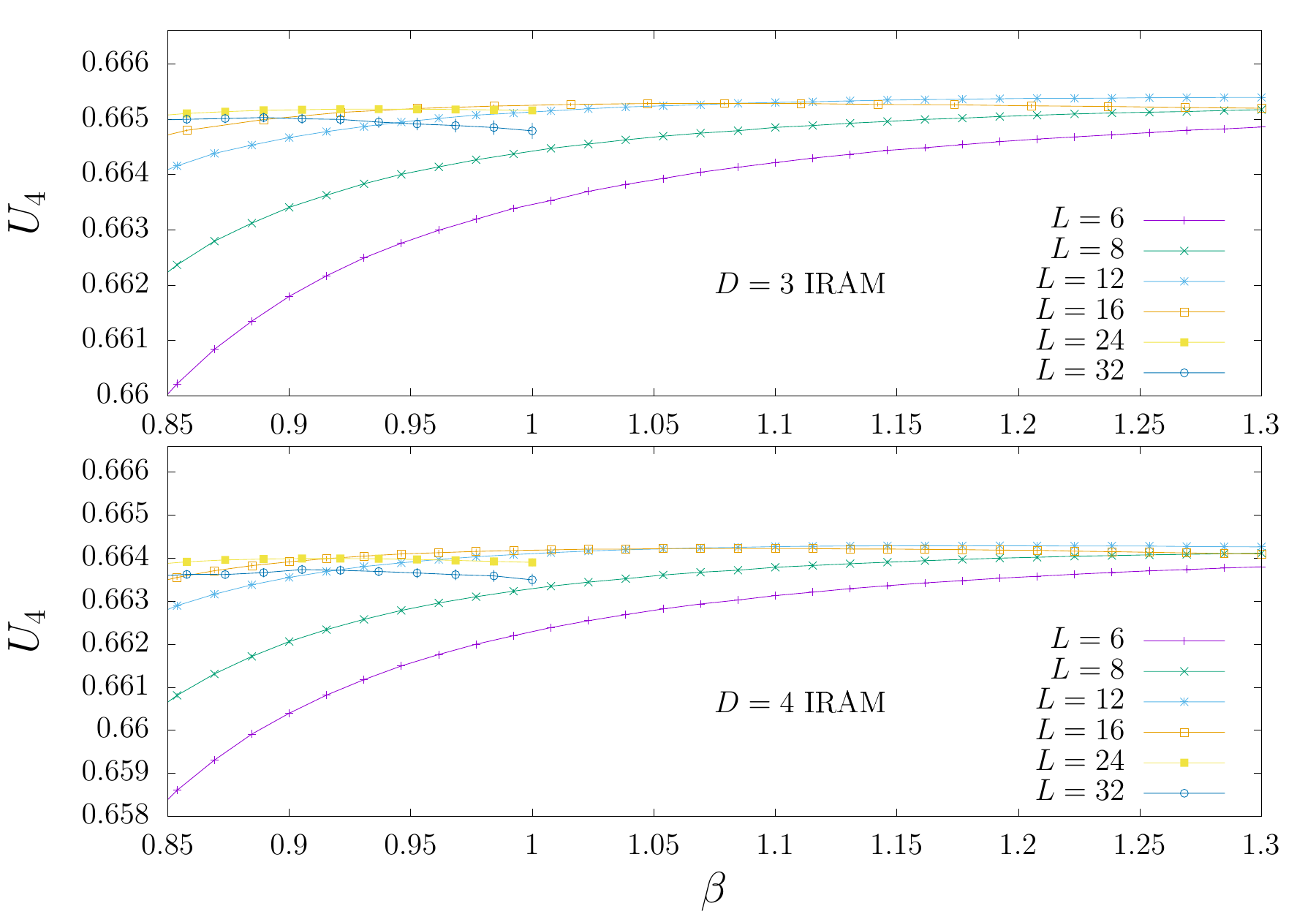}
\caption{(color online)Binder cumulant ($U_4$) as a function of
  the inverse temperature for the isotropic disorder for several
  lattice sizes at $D=3$ (top) and $D=4$ (bottom) in the low-temperature region. Notice the crossing of the different lattice
  sizes and the non monotonic behavior (decrease with $L$) for the
  larger lattices.}
\label{fig:BINDERIRAMEXTD34}
\end{figure}

$R_\xi$ behaves in the low-temperature region as in the high
temperature region: $R_\xi$ decreases as $L$ increases, see Fig.
\ref{fig:XILIRAMEXTD34}, in a completely different way as in a
ferromagnetic phase.

The behavior of the Binder cumulant (see
Fig. \ref{fig:BINDERIRAMEXTD34}) provides the same information. One
can observe the crossing of the $U_4$ curves corresponding to
different lattice sizes and that, for large lattice sizes and a fixed
lower temperature, $g_4$ decreases with $L$, just as in a phase with a
Gaussian distribution (with zero mean) of the magnetization.

\begin{figure}[hbt!]
\centering
\includegraphics[width=\columnwidth, angle=0]{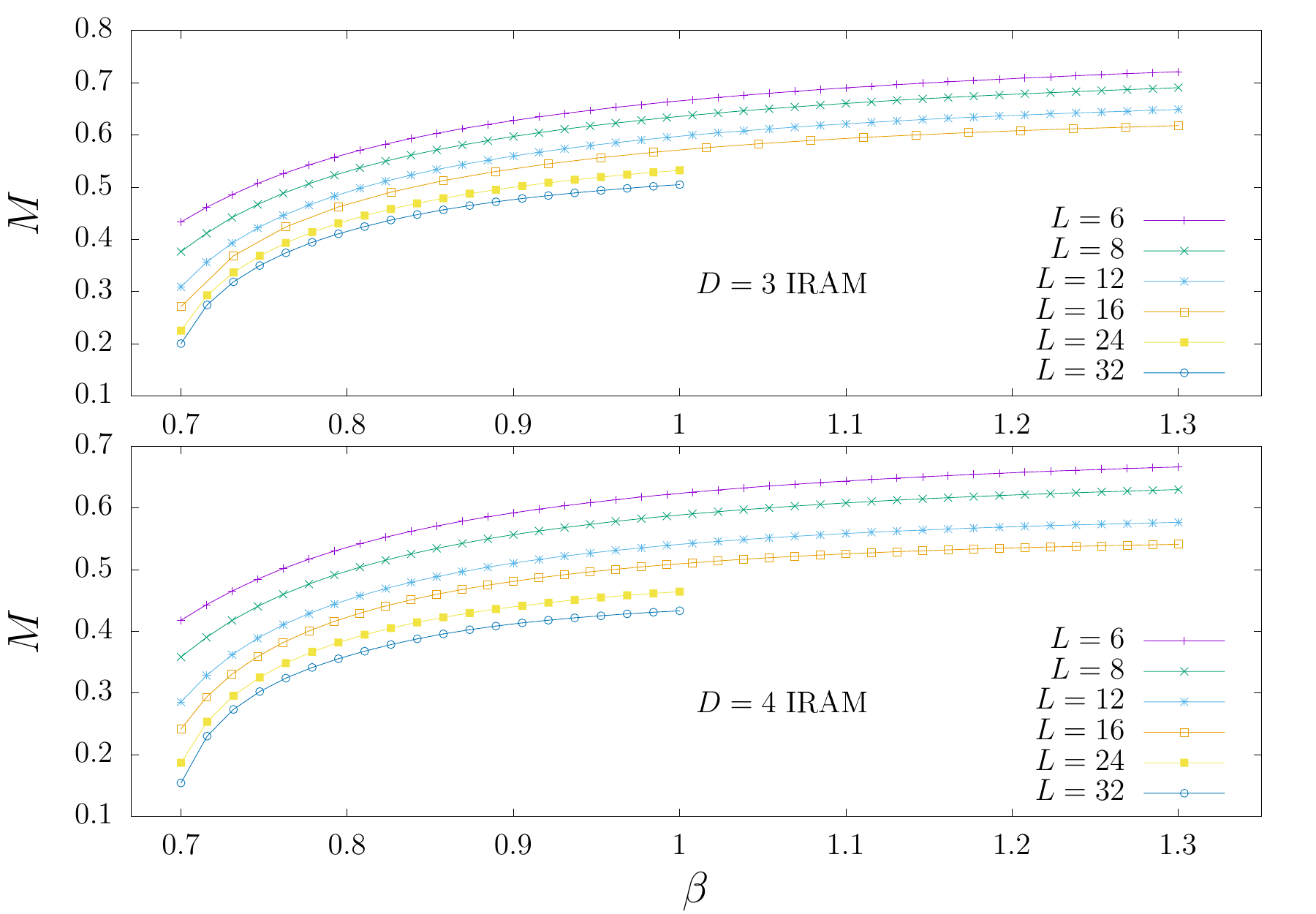}
\caption{(color online) Modulus of the magnetization, $M$, as a
  function of the inverse temperature for the isotropic disorder for
  several lattice sizes at $D=3$ (top) and $D=4$ (bottom).}
\label{fig:MAGIRAMEXTD34}
\end{figure}

In order to understand what are the properties of the low-temperature
phase (we have termed as ZM) we have computed the modulus of the
magnetization.  The behavior of the modulus of the magnetization is
presented in Fig. \ref{fig:MAGIRAMEXTD34} for the two values of the
disorder strength.  Notice that the behavior of these
observables is completely different from the anisotropic case, see
Fig. \ref{fig:XILARAMEXT}-bottom, In the latter case, one can see that
the magnetization is already nonzero and independent of $L$ for relatively
small lattice sizes. However in the isotropic case, we observe that
the magnetization is not asymptotic and decreasing for $D=3$ and 4 and
for all the range of simulated temperatures.

 To be more quantitative, we have analyzed the behavior of $M$ as a
 function of $L$ at $\beta=1$ and we have obtained that $M\sim
 L^{-0.16}$ and $L^{-0.22}$ for $D=3$ and 4, respectively. In
 addition, $R_\xi$ at this $\beta=1$ also decreases as $R_\xi\sim
 L^{-0.16}$ and $L^{-0.10}$ (for the larger sizes) for $D=3$ and 4, respectively. Thus, $M$
 and $R_\xi$ seem to go to zero as the lattice size increases.

  To support this conclusion, we have explicitly checked in Appendix
  \ref{MAG} that at $\beta\simeq 2\beta_c^1\sim 1.4$ and for $D=4$ the
  magnetization and $R_\xi$ decrease with $L$ and are well described
  by power laws (with the exponents similar to those obtained at
  $\beta=1$), meaning that they will vanish in the thermodynamic
  limit. This is a strong hint that this low-temperature phase has
  zero magnetization and finite magnetic correlation length ($\xi$).

From the above observations, one can describe a strong crossover in
the behavior of the isotropic disorder: for $L\gg L_c$ there is no
phase transition in the magnetization channel. We are tempting to
interpret this crossover length ($L_c$) as that proposed by
Imry-Ma. Our power law fits suggest that $L_c$ is much higher than the
maximum simulated lattice size (i.e., $L=48$) for both values of the
anisotropy strength near $\beta_c^1$.  Finally, in
the next section we will argue in favor of the appearance of a phase
with spin-glass properties as an additional argument stating the
magnetization channel is not critical.\footnote{In the following we
will use the word ``transition'' with quotation marks to refer to one of these two crossing
regions which eventually will merge.}

The main conclusion of this subsection and Appendix \ref{MAG} is that the
low-temperature phase presents zero magnetization and finite
correlation length.

However, the numerical results presented in this subsection cannot
completely rule out that the difference $\beta_c^1(L, 2L)-\beta_c^2(L,
2L)$ remains finite or even the magnetization is different from zero
(asymptotically), although we consider this scenario unlikely since it
goes against the Imry-Ma argument.

The behavior of $R_\xi$ around the first phase ``transition''
($\beta_c^1$) looks similar to that of a QLRO one due to the overall
behavior of $R_\xi$: firstly grows (due to $\beta_c^1$) and then
decreases (due to $\beta_c^2$) inducing an apparent merging of the
$R_\xi$ in the interval $[\beta_c^1, \beta_c^2]$ which recalls the
behavior in a QLRO phase transition.  In a QLRO phase the curves of
$R_\xi=\xi/L$ merge because the correlation length is divergent in the
whole low-temperature phase.  See Appendix \ref{XY} for the behavior
of $R_\xi$ and the Binder cumulant of the two-dimensional XY model
which undergoes this QLRO phase transition.

\begin{table*}[hbt!]
\caption{Quotient method results for $D=3$ and $D=4$ and isotropic
  disorder distribution (IRAM) from the two crossing points of $R_\xi$
  for lattice sizes $L_1$ and $L_2$.}  \centering
\label{tab:cross_merg}
\begin{tabular}{|c|c|c|c|c|}
  \hline
  $D$ & $L_1/L_2$  & $\beta_c^1$ & $\beta_c^2$ & $\beta_c^2-\beta_c^1$\\ \hline\hline
3&6/12 & 0.6975(2) & 1.188(1) & 0.491(1)\\
3&8/16 & 0.6977(2) & 1.034(2)  &0.336(2) \\
3&12/24& 0.6978(2) & 0.860(3) & 0.162(3)\\
3&16/32& 0.6980(2) & 0.805(5) & 0.107(5)\\
\hline\hline
4&6/12 & 0.7060(3) & 1.073(1) & 0.367(1)\\
4&8/16 & 0.7065(3) & 0.928(2) & 0.222(2)\\
4&12/24& 0.7055(2) & 0.831(2) & 0.125(2)\\
4&16/32& 0.7052(2) & 0.786(4) & 0.081(4)\\ \hline\hline
\end{tabular}
\end{table*}

\subsubsection{Overlap channel}

In this subsection we will characterize the properties of the ZM low-temperature phase found in the previous subsection, the first
candidate will be a phase with spin-glass properties: it presents zero
magnetization and it is not critical in the magnetic channel. The
presence of a low-temperature phase with spin-glass properties has been previously reported (e.g., see Refs. \cite{Itakura03,Billoni05}).

To this end, we will analyze the behavior of $R_{\xi_q}$ defined
using the overlap correlation length, see Eq. (\ref{XIq}).
To proceed further, we have simulated two real replicas (the same
disorder) of the model and computed the overlap, Eq. (\ref{overlap}),
which characterizes a spin-glass phase. Additionally, we have computed the
associated $R_{\xi_q}$.

\begin{figure}[hbt!]
\centering
\includegraphics[width=\columnwidth, angle=0]{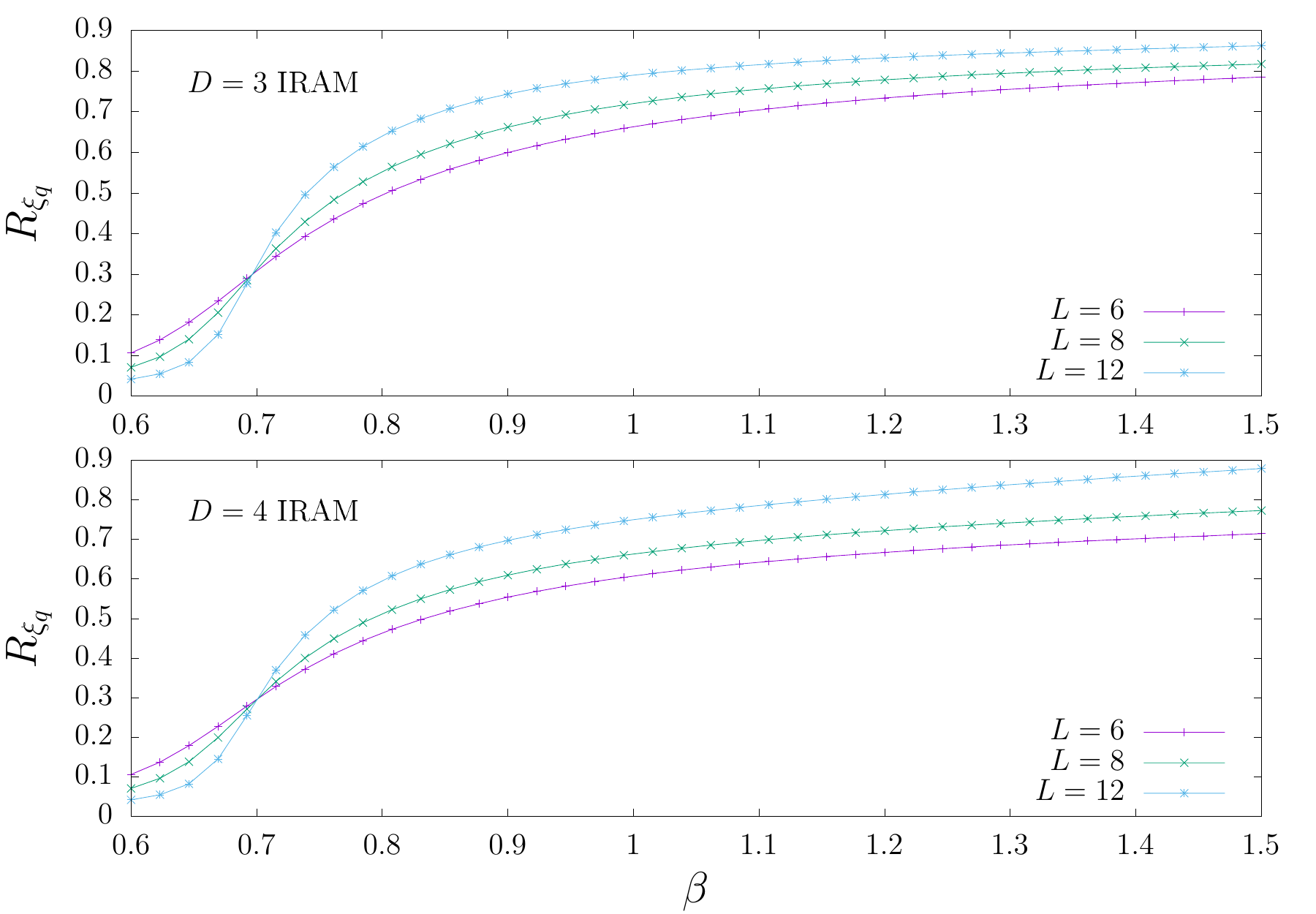}
\caption{(color online) $R_{\xi_q}$ cumulant (overlap channel) as a function of the inverse temperature for the isotropic disorder for several lattice sizes for 
$D=3$ (top) and $D=4$ (bottom).}
\label{fig:XILIRAMSGD4}
\end{figure}

In Fig.~\ref{fig:XILIRAMSGD4} we show the behavior of $R_{\xi_q}$ in a
wide range of inverse temperatures for the  simulated
lattice sizes: the curves cross in the region near $\beta\sim 0.70$.

As we have discussed in the previous
subsection (see also Appendix \ref{MAG}), the spontaneous
magnetization vanishes in the very low-temperature region: in this
region $R_{\xi_q}=\xi_q/L$ diverges as corresponds to the spin-glass
phase.  Notice that $\xi_q$ is a true correlation length in the
paramagnetic phase. In the spin-glass phase $R_{\xi_q}$ diverges as
$L$ increases \cite{ballesteros:98, amit}.

We would like to remark that,  for instance at  $\beta=1.3$ (and for both values of $D$), $R_{\xi}$ 
decreases monotonically with $L$ (magnetic channel)  and $R_{\xi_q}$ increases monotonically with $L$ (overlap channel), for all the simulated values of $L$. 

In Appendix \ref{EA} we show the behavior of the three-dimensional
Edwards-Anderson model, a spin glass, in both the magnetic and the
overlap channels. In this case, the system presents a clear crossing
of the different $R_{\xi_q}$ curves and no signal in the Binder and
$R_\xi$ in the magnetic channel (in this case both observables are
just compatible with zero, showing no critical behavior).

Summing up, the results shown in Fig.~\ref{fig:XILIRAMSGD4} point to a
low-temperature phase with the properties of a spin
glass.\footnote{The system could present a complicated
anti-ferromagnetic low-temperature order inducing the reported
phenomenology of the isotropic disorder~\cite{sgbook}, but, we
consider this scenario very unlikely.}

Hence, the analysis in the overlap channel gives additional weight to
the observations made for the magnetization channel: the PM-FM and the
FM-PM ``transitions'' merge for large $L$ and the final outcome is the
absence of a FM phase for all positive temperatures. Notice that the
vanishing of $R_\xi$ shows that the magnetic correlation length is
finite and this phase is not critical in the magnetization channel. In
particular this result ruled out the QLRO scenario.
 
\begin{figure}[hbt!]
\centering
\includegraphics[width=\columnwidth, angle=0]{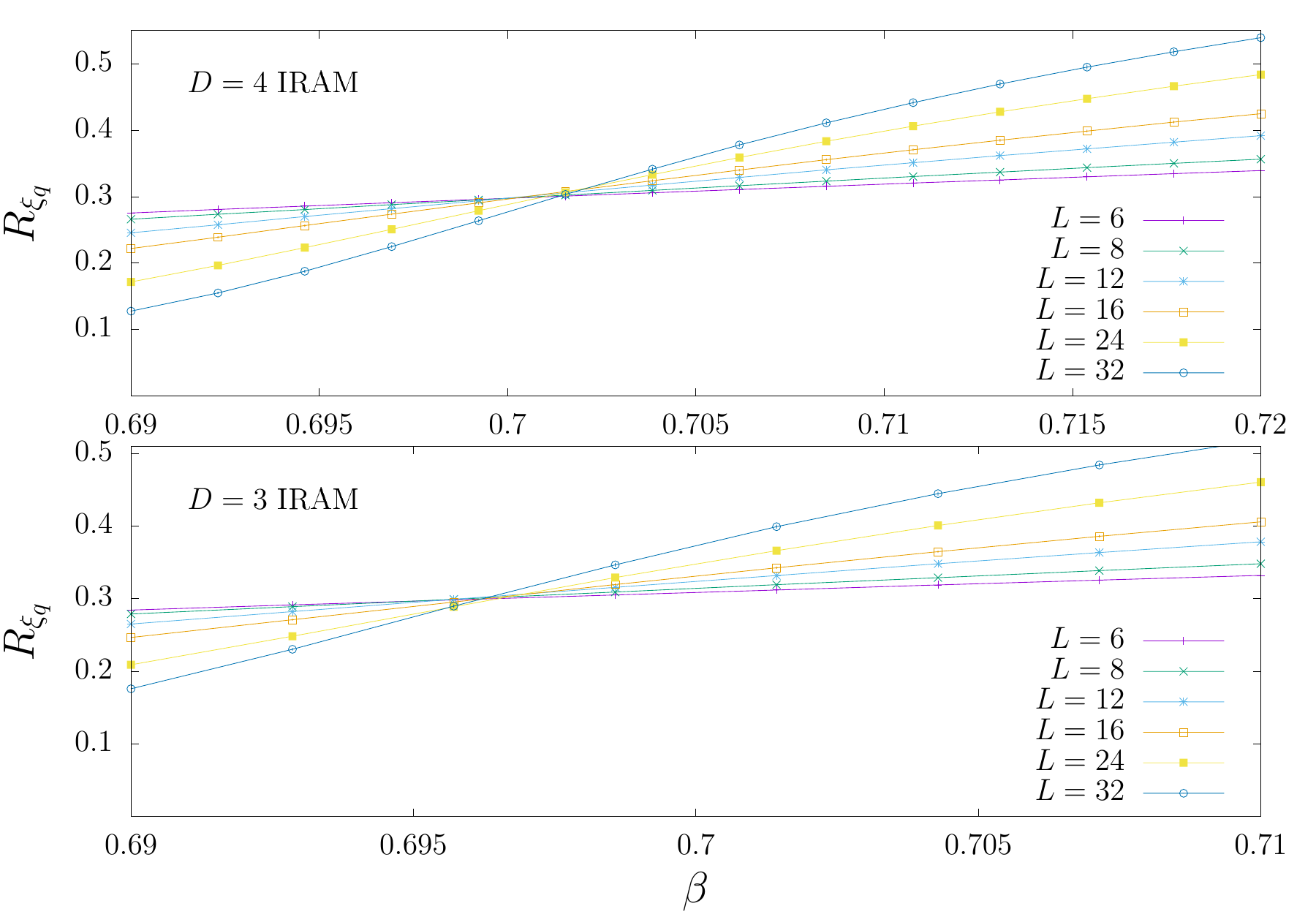}
\caption{(color online) Near the PM-SG phase transition: 
$R_{\xi_q}$ cumulant (overlap channel) as a function of the inverse 
temperature for the isotropic disorder for several lattice sizes at $D=4$ 
(top) and $D=3$ (bottom).}
\label{fig:XILIRAMSGD34}
\end{figure}
  
We have also performed a detailed study of the crossing region of
$R_{\xi_q}$ for both values of $D$. In Fig. \ref{fig:XILIRAMSGD34} we
show the behavior of $R_{\xi_q}$ in the crossing region.
The corresponding crossing points and associated critical
exponents (computed with the quotient method) are presented in Table
\ref{tab:resultsQI}.

\begin{table*}[hbt!]
\caption{Quotient method results for  $D=3$ and $D=4$ and  isotropic
  disorder distribution  (IRAM) from the crossing points of $R_{\xi_q}$ (SG channel) for lattice sizes $L_1$
  and $L_2$.}  \centering
\label{tab:resultsQI}
\begin{tabular}{|c|c|c|c|c|c|}
\hline
$D$ & $L_1/L_2$  & $\beta_\text{cross}$ & $R_{\xi_q}$ & $\nu_\xi$ &
  $\eta_q$  \\
\hline\hline
3 &6/12&  0.6955(3)  & 0.298(1) & 0.776(4) & 1.03(1) \\
3 &8/16 & 0.6965(3)  & 0.302(1) & 0.789(4) & 1.04(1) \\
3 &12/24& 0.6969(2)  & 0.306(2) & 0.797(5) & 1.04(2) \\
3 &16/32& 0.6962(2)  & 0.300(2) & 0.778(6) & 1.06(2) \\
\hline\hline
4 &6/12 & 0.6999(2)  & 0.2973(8)& 0.799(3) & 1.02(1) \\
4 &8/16 & 0.7003(2)  & 0.299(1) & 0.804(4) & 1.03(1) \\
4 &12/24& 0.7014(2)  & 0.305(2) & 0.822(6) & 1.02(2) \\
4 &16/32& 0.7020(2)  & 0.311(2) & 0.818(7) & 1.00(2) \\
\hline\hline
\end{tabular}
\end{table*}

Our results point to (see Table \ref{tab:resultsQI}) $\nu\sim 0.78-0.82$ for both
values of $D$. In addition, we do not see the growing of the $\nu$ exponent with $L$ as seen in the magnetization channel. 
In addition, the computed exponent is very different from that reported for Ising and Heisenberg spin glass models, 
see  Table \ref{tab:ModelsSG}.

However, we have for  $\eta$ a big value, near 1. The overlap $q$ and the
magnetization $m$ have different scaling dimensions, $d_q$ and $d_m$,
correspondingly.  Roughly $q \sim m^2$, so that $d_q\simeq
2d_m$. Substituting in this estimate $2 d_m=d-2+\eta$ and $d_q= (d-2+
\eta_q)/2$ one gets for $d=3$ and $\eta(d=3) \simeq 0$ (see Table
\ref{tab:resultsI} of Appendix \ref{ISO}): $\eta_q=1+2 \eta \simeq 1$ in a very good
agreement with the $ \eta_q$ values reported in Table
\ref{tab:resultsQI}. This is the consequence of simulating lattice sizes smaller than the (Imry-Ma) crossover length
in the critical region.
From previous discussion, the value of this crossover length at very low temperatures is small, allowing us to monitor the decreasing of the magnetization and $R_\xi$ for the lattice sizes we have been able to simulate.

Therefore, the exponents of Table \ref{tab:resultsQI} can be
considered only as effective ones.  The simplest explanation is that
they first attain the PM-FM values (in a ferromagnetic phase $q\propto m^2$)  and eventually they will cross over
to the PM-SG ones in the asymptotic limit (where the magnetization is zero with a nonvanishing overlap), since we are simulating
lattice sizes below the Imry-Ma length.

\begin{table*}[hbt!]
\caption{Critical exponents for two related three-dimensional
  models  undergoing PM-SG phase transitions.}  \centering

\label{tab:ModelsSG}
\begin{tabular}{|c|c|c|c|}
\hline
Model & $\nu$ & $\eta_q$ \\
\hline\hline
Heisenberg~\cite{fernandez:09} & 1.5(2) & -0.19(2)\\
Edwards-Anderson (Ising)~\cite{janus:13}   &  2.56(4) & -0.390(4)\\ \hline \hline
\end{tabular}
\end{table*}

\begin{table*}[hbt!]
\caption{Critical exponents and cumulants for three three-dimensional
  related models  undergoing PM-FM phase transitions.}  \centering

\label{tab:Models}
\begin{tabular}{|c|c|c|c|c|c|c|}
\hline
Model & $\nu$ & $\eta$ & $\omega$ & $R_\xi$ & $U_4$ & $g_2$\\
\hline\hline
Heisenberg~\cite{campostrini:02,hasenbusch:11,hasenbusch:01} &  0.7116(10)   & 0.0378(3) & 0.773            &0.5639(2) & 0.6202(1)&  0\\
Ising (site-diluted)~\cite{ballesteros:98}        &  0.6837(53)   & 0.037(4)      & 0.37(6)        & 0.598(4) & 0.449(6)   & 0.145(3)\\ 
Ising~\cite{ferrenberg:18,simmons:17}     &  0.629912(86) & 0.0362978(20) & 0.8303(18)     &0.6431(1)          & 0.46548(5) & 0 \\ \hline \hline
\end{tabular}
\end{table*}

\section{Conclusions and Discussions} \label{V}

We have simulated the RAM with cubic and isotropic disorder in three
dimensions. To do that, we have run very large numerical simulations
using PT (and Metropolis as the local update method) simulating large
lattice sizes ($L=48$).

This model was also studied in Ref.~\cite{Nguyen09} by means numerical
simulations and using the PT algorithm for cubic and isotropic
disorder distributions near $\beta\sim 0.7$. The authors simulated
only $D=4$ and $L \le 24$. They found clear crossing of the different
Binder cumulant curves for both distributions. Despite this, the
authors claimed that the low-temperature phase is in the QLRO
class. They computed only the $\nu$ exponent by studying the
susceptibility, quoting for both disorders $\nu=0.70998$ with 0 as a
statistical error, hence, claiming that both disorders belong to the
same universality class. In addition they reported
$\beta_c=0.70998(4)$ for ARAM and $\beta_c=0.70435(2)$ for IRAM. In
this work, we report estimates of $\beta_c$ and $\nu$ for $D=4$ not
compatible with those of Ref.~\cite{Nguyen09} for the anisotropic
disorder. In addition for isotropic disorder, we have characterized
how the two phase transitions eventually merge and the low-temperature
phase can be consistently described by a phase with spin-glass properties.

In the previous section, we have discussed the critical behavior
for both disorder distributions. Since it is clear that the critical
behavior of these two  distributions is different we will discuss
separately our findings for each distribution.

\subsection{Cubic disorder}

We have found a phase transition between a PM phase and a FM one. Both
values of the strength of the anisotropy, $D$, provide with critical
exponents and cumulants compatible in the statistical error, fact
which supports the universality of this disorder for moderate values
of $D$.

Notice that our estimates for the critical exponents and cumulants are
completely different from the Heisenberg model ones (see Table
\ref{tab:Models}). Moreover, we have simulated the pure Heisenberg
model within the PT method and used the same procedure of analysis as
used in this paper.  The obvious conclusion is that ARAM and Heisenberg
models belong to different universality classes, cf. Table
\ref{tab:resultsPureXi} of Appendix \ref{O3}.

Furthermore our results do not agree with the perturbative RG
predictions~\cite{Dudka01c,new_Ital,Dudka01b,Mukamel82,Korzhenevskii88,Shapoval}
which state that the ARAM should belong to the same universality class
as the three-dimensional site-diluted Ising model, as follows from
comparison of our values for the critical exponents from
Tables~\ref{tab:extrapolationQA} and \ref{tab:extrapolationFCAA} with
the Monte Carlo results of Table~\ref{tab:Models} for the
three-dimensional site-diluted Ising model. The latter are
corroborated also by RG studies, see e.~g.
Refs.~\cite{Pelissetto02,Folk03,Kompaniets21}.
 
The following scenario emerges from our simulations: the anisotropic
disorder, as predicted by RG, is relevant and changes the universality
class of the pure model. We have checked that this scenario holds for
$D\le 4$. We expect this new fixed point should be relevant for
$0<D<D_c$. We can try to conjecture the behavior of the model with
anisotropic disorder for strong anisotropy.  For large $D>D_c$ the
anisotropic disorder will destroy the FM phase and a SG phase will
arise. We know that for $D=\infty$ the system is described by the
Ising spin-glass universality class~\cite{ParisienToldin06,Liers07}. Open
problems are the characterization of $D_c$ and to figure out the
universality class for $D_c<D<\infty$: is that of the Ising spin-glass
model in three-dimensions?

\subsection{Isotropic disorder}

We have found two different crossing regions of the cumulant $R_\xi$ (and of the $U_4$ cumulant).
Furthermore, these two regions seem to merge, a clear indication of a
phase with zero magnetization for all the positive temperatures. We
have checked this scenario by studying the behavior of the
magnetization and $R_\xi$ at very low temperatures. This result is in
full agreement with the Imry-Ma arguments that predicts zero
magnetization for $T>0$.  Note that Imry-Ma domain arguments were also
corroborated recently by extensive Monte Carlo simulations for
two-dimensional random anisotropy magnets \cite{Garanin22}.

Studying the overlap channel we have found a phase transition
between a paramagnetic phase and a phase with spin-glass properties,
and we have characterized its critical exponents.  However, taking
into account that the (Imry-Ma) crossover length around the transition is much
larger than the largest simulated lattice size we consider these
exponents as effective ones.

A low-temperature phase with spin-glass properties was reported in
Ref. \cite{Billoni05} studying the isotropic disorder model simulating (out-of-equilibrium)
a $L=20$ lattice for $D=3.5$. They found that the
field-cooled and zero-field-cooled magnetization behave, in the low-temperature region, in the same
way as in a spin-glass~\cite{sgbook}.

Our work provides additional arguments (from equilibrium numerical
simulations) supporting a low-temperature phase for the isotropic
disorder with spin-glass behavior and zero magnetization.

An important open problem is to try to compute the asymptotic
exponents of this PM-SG transition and to study how the model leads to
the Ising Edwards-Anderson universality class for
$D=\infty$~\cite{ParisienToldin06,Liers07}.  We consider that such study is
outside of the current available computational resources.

\acknowledgments

The authors acknowledge useful discussions with R. Folk,
A. A. Fedorenko, L. A. Fernandez, V. Martin-Mayor. We also thank the
anonymous Referee for useful suggestions.  This work was partially
supported by Ministerio de Econom\'{\i}a y Competitividad (Spain)
through Grants No.\ FIS2016-76359-P and PID2020-112936GB-I00, by Junta
de Extremadura (Spain) through Grant No.\ GRU18079, GR21014, IB16013
and IB20079 (partially funded by FEDER), Polish National Agency for
Academic Exchange (NAWA) through the Grant No. PPN/ULM/2019/1/00160,
National Academy of Sciences of Ukraine within the framework of the
Project K$\Pi$KBK 6541030, and European Union through Grant
No.\ PIRSES-GA-2011-295302. We have run the simulations in the
computing facilities of the Instituto de Biocomputaci\'{o}n y
F\'{\i}sica de Sistemas Complejos (BIFI) and those of the Instituto de
Computaci\'{o}n Cient\'{\i}fica Avanzada (ICCAEx).  Yu.H. acknowledges
support of the JESH mobility program of the Austrian Academy of
Sciences and hospitality of the Complexity Science Hub Vienna when
finalizing this paper.

\appendix
\section{Numerical Simulations details}
\label{num}

We have performed extensive Monte Carlo numerical simulations using
the Metropolis (with 10 hits) and PT algorithms,
see Refs.~\cite{PT1996,Marinari1998}. We have checked that the PT is correctly working with our choice of
the different parameters. 

In order to have an additional test of the thermalization of our
systems, we have studied the behavior of different observables at all
the temperatures as a function of $\log t$ ($t$ being the Monte Carlo
time). We consider that we have thermalized the system, for a given
observable, when the last three points are compatible in the error
bars and a plateau can be defined (the last point is computed with the
last half of the Monte Carlo history).  All the results presented in
this study fulfill this thermalization criteria.

We will provide the parameters of the different numerical simulations
for the models with $D=3$ and 4 and both types of disorder, cubic and
isotropic.

\begin{itemize}
\item Anisotropic Disorder.
\begin{enumerate}
    \item Near the critical point. See Table \ref{tab:parametersA}.
    \item Extended run.  We have simulated $L=6$, 8 and 12 for $D=4$
      using the PT method: 1000 samples, 40 temperatures and 204800
      Monte Carlo steps for each temperature.
\end{enumerate}
    \item Isotropic Disorder
\begin{enumerate}
\item Near the critical point (Magnetization, first phase
  ``transition"). See Table \ref{tab:parametersI}.

        \item Extended runs (Magnetization). For $D=3$, 1000 samples
          and 40 temperatures (PT) have been simulated
          for $L=6$, 8, 12 and 16; 400
          samples and 40 temperatures (PT) for $L=24$ and 230 samples
          and 20 temperatures (PT) for $L=32$.  We have performed
          409600 sweeps per temperature in the PT.
        
        For $D=4$, 40 temperatures (PT) have been simulated with 2000,
        1200, 2000, 658 and 786 samples for $L=6$, 8, 12, 16 and 24
        respectively; 500 samples and 20 temperatures (PT) for
        $L=32$. We have performed 409600 sweeps per temperature in the
        PT.
        
        \item Extended runs (Overlap). 40 temperatures (PT) have been
          simulated for $L=6$, 8, 12. For $D=3$, we have run 1000
          samples and for $D=4$, 500 samples. We have performed
          204800/409600 sweeps per temperature in the PT for $D=3/4$
          respectively.
        
        \item Near the critical point (overlap). For $D=3$, we have
          simulated 8 temperatures (PT) for $L=6$, 8, 12, 16, 24 and
          32 with 1000, 1000, 1000, 760, 500 and 500 samples
          respectively. For $D=4$, we have simulated 14 temperatures
          (PT) for $L=6$, 8, 12, 16, 24 and 32 with 3000, 3000, 1792,
          1000, 500 and 495 samples respectively. In all cases we have
          performed 204800 sweeps per temperature in the PT.
\end{enumerate}
\end{itemize}

\begin{table}[ht!]
\caption{Parameters used in the numerical simulations for the cubic
  disorder distribution (ARAM). $N_{\rm samples}$ is the number of
  disorder realizations, $N_{\rm T}$ is number of temperatures in the
  parallel tempering and $N_{\rm sweeps}$ number of Monte Carlo sweeps
  per temperature.}  \centering
\label{tab:parametersA}
\begin{tabular}{|c|c|c|c|c|c|}
\hline
$D$  &$L$ & $N_\text{samples}$ & $N_\text{T}$&  $N_\text{sweeps}$\\
\hline
3  & 6 & 13600 & 14 & $10^5$\\
3  & 8 & 4800 & 14 & $10^5$\\
3  & 12 & 2937 & 14 & $10^5$\\
3  & 16 & 2000 & 14 & $1.28\times 10^5$\\
3  & 24 & 2091 & 14 & $1.28\times 10^5$\\
3  & 32 & 1400 & 8 & $2.56\times 10^5$\\
3  & 48 & 392 & 8 & $5.12\times 10^5$\\ \hline
4  & 6 & 13600 & 14 & $10^5$\\
4  & 8 & 4800 & 14 & $10^5$\\
4  & 12 & 2938 & 14 & $10^5$\\
4  & 16 & 2000 & 14 & $1.28\times 10^5$\\
4  & 24 & 2068 & 14 & $2.56\times 10^5$\\
4  & 32 & 1400 & 8 & $2.56\times 10^5$\\
4  & 48 & 768 & 14 & $5.12\times 10^5$\\
\hline
\hline
\end{tabular}
\end{table}

\begin{table}[ht!]
\caption{Parameters used in the numerical simulations for the
  isotropic disorder distribution (IRAM) near $\beta_c^1$.}
\centering
\label{tab:parametersI}
\begin{tabular}{|c|c|c|c|c|c|}
\hline
$D$  &$L$ & $N_\text{samples}$ & $N_\text{T}$&  $N_\text{sweeps}$\\
\hline
3  & 6 & 13600 & 14 & $10^5$\\
3  & 8 & 4800 & 14 & $10^5$\\
3  & 12 & 3000 & 14 & $10^5$\\
3  & 16 & 2000 & 14 & $1.28\times 10^5$\\
3  & 24 & 2075 & 14 & $1.28\times 10^5$\\
3  & 32 & 1400 & 8 & $2.56\times 10^5$\\
3  & 48 & 400 & 8 & $5.12\times 10^5$\\ \hline  
4  & 6 & 17600 & 14 & $10^5$\\
4  & 8 & 4800 & 14 & $10^5$\\
4  & 12 & 3000 & 14 & $10^5$\\
4  & 16 & 2000 & 14 & $1.28\times 10^5$\\
4  & 24 & 2000 & 14 & $2.56\times 10^5$\\
4  & 32 & 1400 & 8 & $2.56\times 10^5$\\
4  & 48 & 400 & 8 & $1.024\times 10^6$\\ \hline \hline
\end{tabular}
\end{table}

\section{Quotient and fixed coupling methods}
\label{quotient}
In this appendix we briefly describe the quotient and fixed coupling
methods.

Firstly, we describe the quotient method.  Let $O(\beta,L)$ be a
dimensionful quantity scaling in the thermodynamic limit as
$\xi^{x_O/\nu}$. For a dimensionless observable the exponent
$x_O=0$. Thereafter, we will use the symbol $g$ to denote all the
dimensionless quantities, such as $U_4$, $g_2$ or $R_\xi$.

The behavior of the observable $O$ can be studied by computing, at $L$
and $2L$, the quotient
\begin{equation}
{\mathcal
  Q}_O=\frac{O(\beta_\text{cross}(L, 2L),2L)}{O(\beta_\text{cross}(L, 2L),L)}\,,
  \end{equation}
  where $\beta_\text{cross}(L, 2L)$ is defined by
  \begin{equation}
g(L,\beta_\text{cross}(L, 2L))=g(2L, \beta_\text{cross}(L, 2L))\,.
  \end{equation}
  
From the previous discussion, one can write
\begin{equation}\label{eq:QO}
{\mathcal Q}_O^{\,\mathrm{cross}}=2^{x_O/\nu}+\mathcal{O}(L^{-\omega})\,,\
\end{equation}
and 
\begin{equation}\label{eq:QOB}
g^{\,\mathrm{cross}}=g^* +\mathcal{O}(L^{-\omega})\,,\
\end{equation}
where $x_O/\nu$, $g^*$ and the correction-to-scaling exponent $\omega$
are universal quantities. In order to compute the $\nu$ and $\eta$
exponents, one should study dimensionful observables such as the
susceptibility ($x_\chi= \nu(2-\eta)$) and the $\beta$-derivatives of
$R_\xi$ and $U_4$ ($x=1$ in both cases).

The crossing point of the inverse
temperature ($\beta_\text{cross}(L,2L)$) behaves following the equation
\begin{equation}\label{eq:Tc}
\beta_\text{cross}(L,2L) = \beta_\text{c} + A_{\beta_\text{c},g} L^{-\omega -
  1/\nu}+\ldots\,.
\end{equation}

The leading correction-to-scaling exponent can be computed via 
the quotient of a given dimensionless quantity $g$ ($Q_g$). The behavior of this  quotient
is
\begin{equation}\label{eq:dimensionless-quotients}
\mathcal Q^{\text{cross}}_g(L) = 1 + A_g L^{-\omega} + B_g L^{-2\omega} + \ldots .
\end{equation}

Another way to compute critical exponents is to work at a fixed
dimensionless observable. One fixes the value of the dimensionless
observable $g=g_f$ near the universal one (for example fixing a given
value of $R_\xi$) and then computes $\beta(L)$ defined as
\begin{equation}
g_f=g(\beta(g_f, L), L)\,.
\end{equation}
Using these values of the inverse temperature, one can monitor the
scaling.  At this value of the inverse temperature we can study the
scaling of different observables (e.g., susceptibility, derivatives of
$R_\xi$ and Binder cumulant, etc) to extract the critical exponents
using
\begin{equation}
O(\beta(g_f,L),L)=A(g_f) L^{x_O/\nu}\left (1 + {\cal O} \left(\frac{1}{L^\omega}\right)\right)\,.
\end{equation}

\section{The PM-FM ``transition'' for the isotropic disorder}
\label{ISO}

We have shown numerical data based on the behavior of $R_\xi$ with
temperature and lattice sizes supporting that both phase ``transitions''
(near $\beta_c^1$ and $\beta_c^2$) are merging, eventually for lattice
sizes bigger than a crossover length as stated by the Imry-Ma
argument.

In this appendix we report the values of the critical exponents of the
PM-FM ``transition" observed for IRAM at temperature $T\sim
1/\beta_c^1$. As we have shown in the main text, this phase ``transition''
should disappear for large lattice sizes. Therefore the reported below
values of the exponents characterize the effective critical behavior
of the model at least for $L \le 48$.  This behavior is observed
before the PM-FM ``transition'' point $\beta_c^1$ merges with FM-PM
transition point $\beta_c^2$ moving from the low-temperature region.

\begin{figure}[hbt!]
\centering
\includegraphics[width=\columnwidth, angle=0]{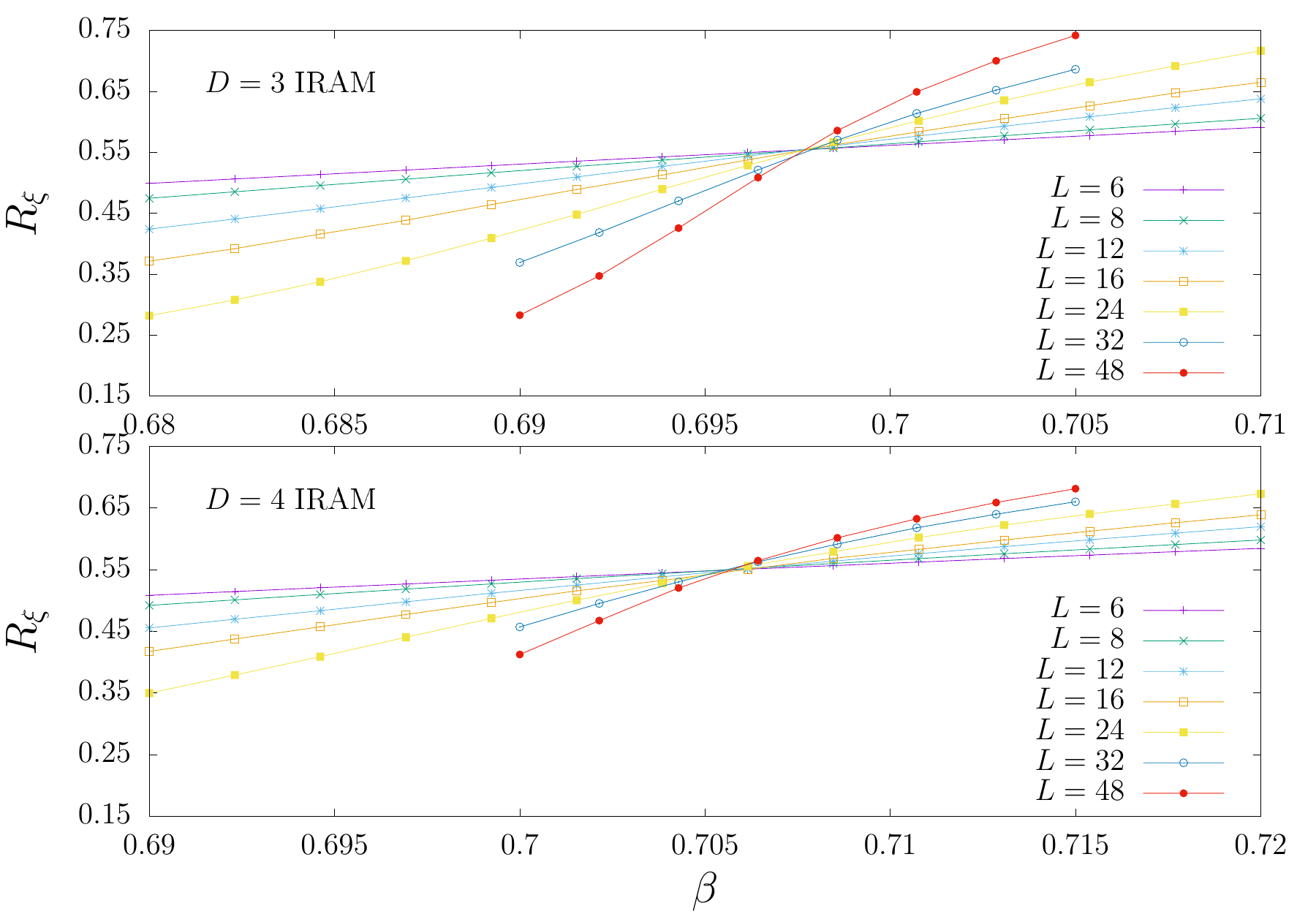}
\caption{(color online) Effective critical behavior in vicinity of
  $\beta_c^1$: $R_\xi$ cumulant as a function of the inverse
  temperature for the isotropic disorder for several lattice sizes.}
\label{fig:XILIRAM}
\end{figure}

\begin{figure}[ht]
\centering
\includegraphics[width=\columnwidth, angle=0]{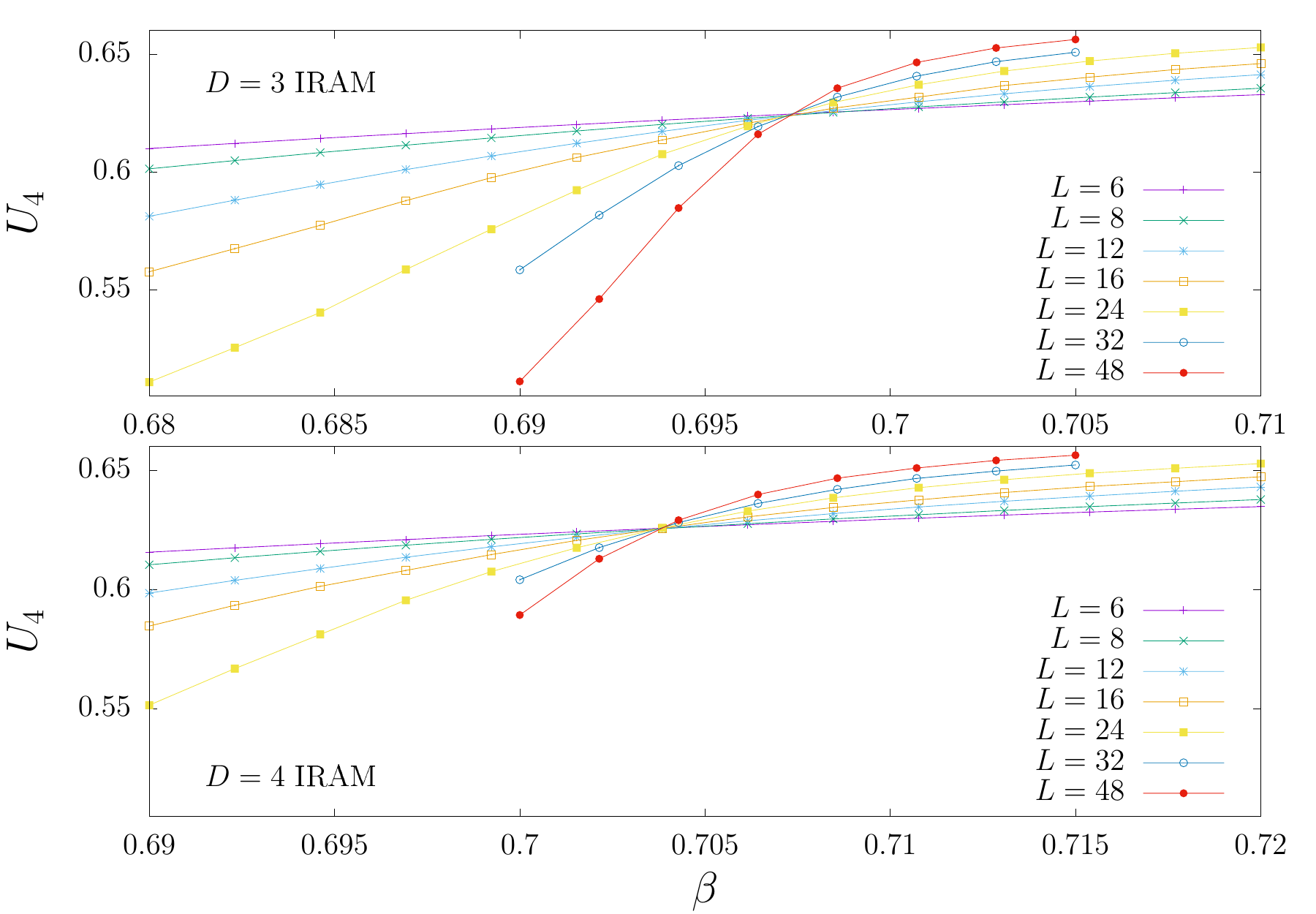}
\caption{(color online) Effective critical behavior in vicinity of $\beta_c^1$: 
Binder cumulant as a function of the inverse
  temperature for the isotropic disorder for several lattice sizes.}
\label{fig:G4IRAM}
\end{figure}

In Figs. \ref{fig:XILIRAM} and \ref{fig:G4IRAM} we show the behavior
of $R_\xi$ and $U_4$ in the vicinity of the first phase transition
temperature $\beta_c^1$.  We have analyzed the isotropic disorder in
the same way as the cubic one, (see section \ref{IVA}), focusing on the
analysis of the crossing temperatures of $R_\xi$ since the Binder
cumulant data present strong corrections to scaling.

We report in Table \ref{tab:resultsI} our results. 
Note that our estimates of the effective critical exponents
and cumulants are different from the pure Heisenberg ones.  Moreover,
the $\nu$ exponent is measurably growing with the disorder strength,
which is the right behavior in view of the onset of a second phase
transition merging with this one.

\begin{table*}[hbt!]
\caption{Quotient method results for  $D=3$ and $D=4$ and  isotropic
  disorder distribution  (IRAM) from the crossing points of $R_\xi$ for lattice sizes $L_1$
  and $L_2$ near $\beta_c^1$.}  \centering

\label{tab:resultsI}
\begin{tabular}{|c|c|c|c|c|c|c|c|c|}
  \hline
  $D$ & $L_1/L_2$  & $\beta_\text{cross}$ & $R_\xi$ & $\nu_\xi$ & $\nu_{U_4}$ &
  $\eta$ & $Q_{U_4}$ & $Q_{g_2}$ \\
\hline\hline

3&6/12 & 0.6975(2) & 0.5540(8)& 0.807(2)  &  0.768(2)& 0.012(4) & 0.9995(1)& 1.64(4)\\
3&8/16 & 0.6977(2) & 0.554(1) & 0.828(6)  &  0.81(1)& 0.015(5) & 1.0008(2)& 1.69(5)\\
3&12/24& 0.6978(2) & 0.556(2) & 0.846(6)  &  0.83(1) & 0.013(6) & 1.0030(2)& 1.40(5)\\
3&16/32& 0.6980(2) & 0.557(2) & 0.87(1) &  0.89(1)& 0.00(1)& 1.0049(4)& 1.18(4)\\
3&24/48& 0.6975(2) & 0.552(3) & 0.92(2) &  0.90(2)  & 0.01(2)& 1.0033(6)& 1.4(1)\\\hline\hline

4&6/12 & 0.7060(3) & 0.551(1) & 0.896(4) & 0.869(4)& -0.004(5) & 1.0024(2) & 1.36(2) \\
4&8/16 & 0.7065(3) & 0.553(2)  & 0.98(1)& 0.97(2) & -0.012(7) & 1.0049(3) & 1.31(4) \\
4&12/24& 0.7055(2) & 0.548(2)  & 0.98(1)& 1.01(2) & -0.007(9) & 1.0048(4) & 1.28(4) \\
4&16/32& 0.7052(2) & 0.542(3)  & 0.97(2)  & 0.98(2)   & -0.00(1) & 1.0049(5) & 1.14(4) \\
4&24/48& 0.7055(3) & 0.549(5)  & 1.13(5)  & 1.17(5)   & -0.04(2)  &1.0077(8) & 1.04(7) \\ 
 \hline\hline

\end{tabular}
\end{table*}

Finally, we recall
that the Imry-Ma characteristic lengths for these values of $D$ is
much bigger than $L=48$.

\section{The magnetization and $R_\xi$ at very low temperature for isotropic disorder with strength $D=4$}
\label{MAG}
  In this appendix we analyze the behavior of the magnetization at
  very low temperature.
  
  To compute the magnetization at $T\simeq T_c^1/2 = 1/(2 \beta_c^1)
  \simeq 0.709$ we have resorted to the use of a out-equilibrium
  dynamical approach consisting in the computation of the magnetization
  and $R_\xi$ as a function of time, starting from ordered and fully
  disordered configurations, and to extrapolate both simulations to
  the same asymptotic value using power laws: $m_\infty+{\cal
    O}(t^{-x_m})$ or $R_\xi=R_{\xi_\infty}+{\cal O}(t^{-x_{R_\xi}})$.
 
We have simulated $L=8$ (400 samples and 2048000 Monte Carlo steps),
$L=16$ (300 samples and 2048000 Monte Carlo steps), $L=24$ (200
samples and 16384000 Monte Carlo steps) and $L=32$ (50 samples and
16384000 Monte Carlo steps).
    
\begin{figure}[ht]
\centering
\includegraphics[width=\columnwidth, angle=0]{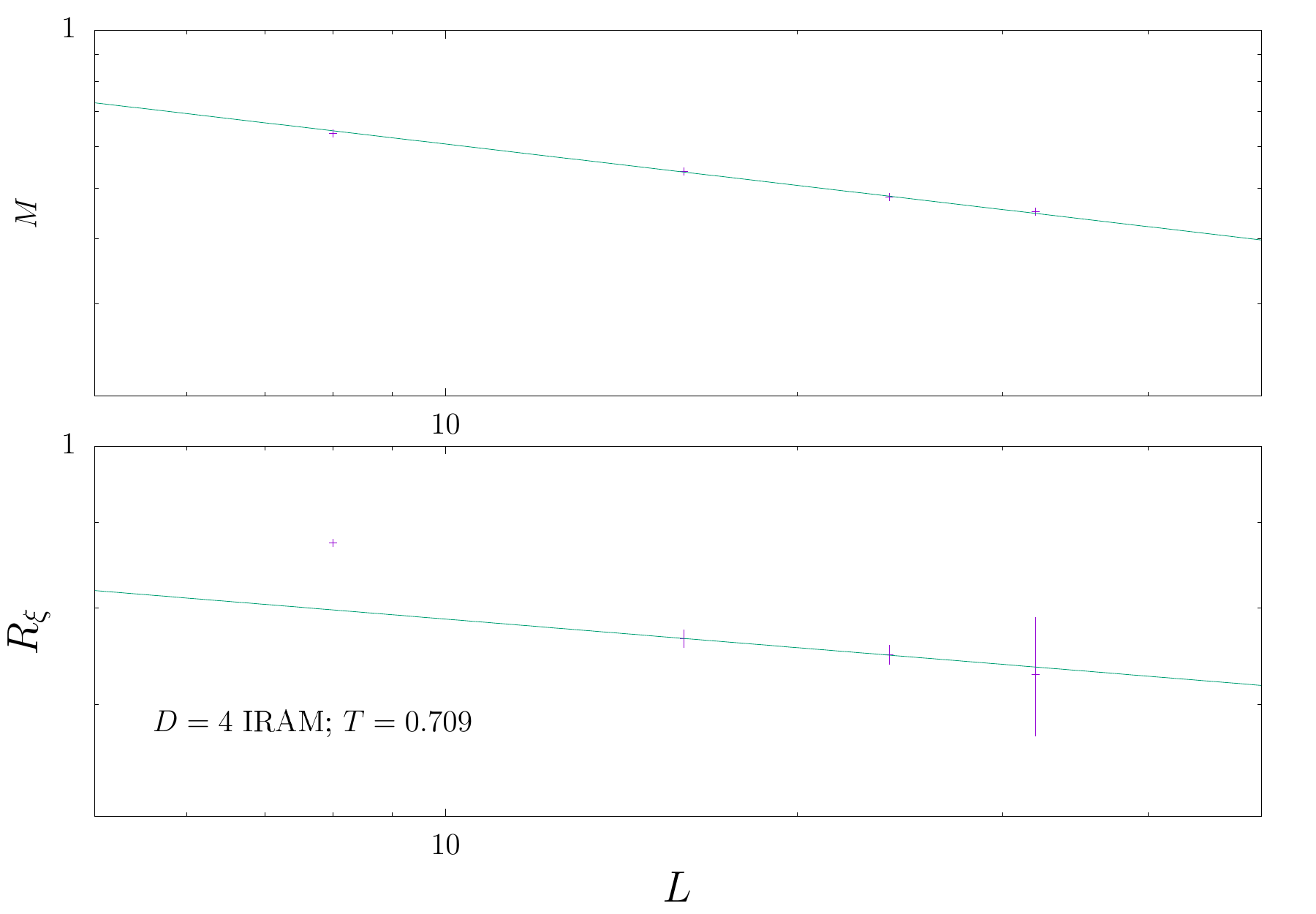}
\caption{(color online) Behavior of the magnetization per spin (top)
  and $R_\xi$ (bottom) for $D=4$ and isotropic disorder as a function
  of the lattice size at $T=0.709$. The data are well described by
  power laws (continuous lines in both panels).}
\label{fig:MXI4IRAM}
\end{figure}

We have found that the data (in the range of the lattice sizes we are
able to simulate at this low temperature) are well described by power
laws: the magnetization scales to zero as $L^{-0.25}$ and $R_\xi$ as
$L^{-0.15}$ (see Fig. \ref{fig:MXI4IRAM}). Notice that in a
noncritical phase, with the thermodynamic correlation length greater
than $L$, the behavior of this cumulant should be proportional to
$L^{-1}$: hence, we are still simulating in the range of lattice sizes
smaller than the asymptotic value of the correlation length.

Therefore, from this analysis, the behavior of the isotropic disorder
at very low temperatures can be described by a phase with zero
magnetization and finite correlation length in the magnetic
channel. This result supports pretty well the behavior of the $R_\xi$
cumulant and the magnetization around and below the second phase ``transition'' at $\beta_c^2$ described in Sec. \ref{IVB}.

\section{The two-dimensional $O(2)$ model}
\label{XY}

In order to test the QLRO scenario for the IRAM, we have simulated the
two-dimensional $O(2)$ model.

The most accurate analysis of this model has been performed in
Ref.~\cite{hasenbusch:05} reporting $\beta_c=1.1199(1)$,
$R_\xi=0.7506912$ and $U_4=0.660603(12)$.

\begin{figure}[hbt!]
\centering
\includegraphics[width=\columnwidth,  angle=0]{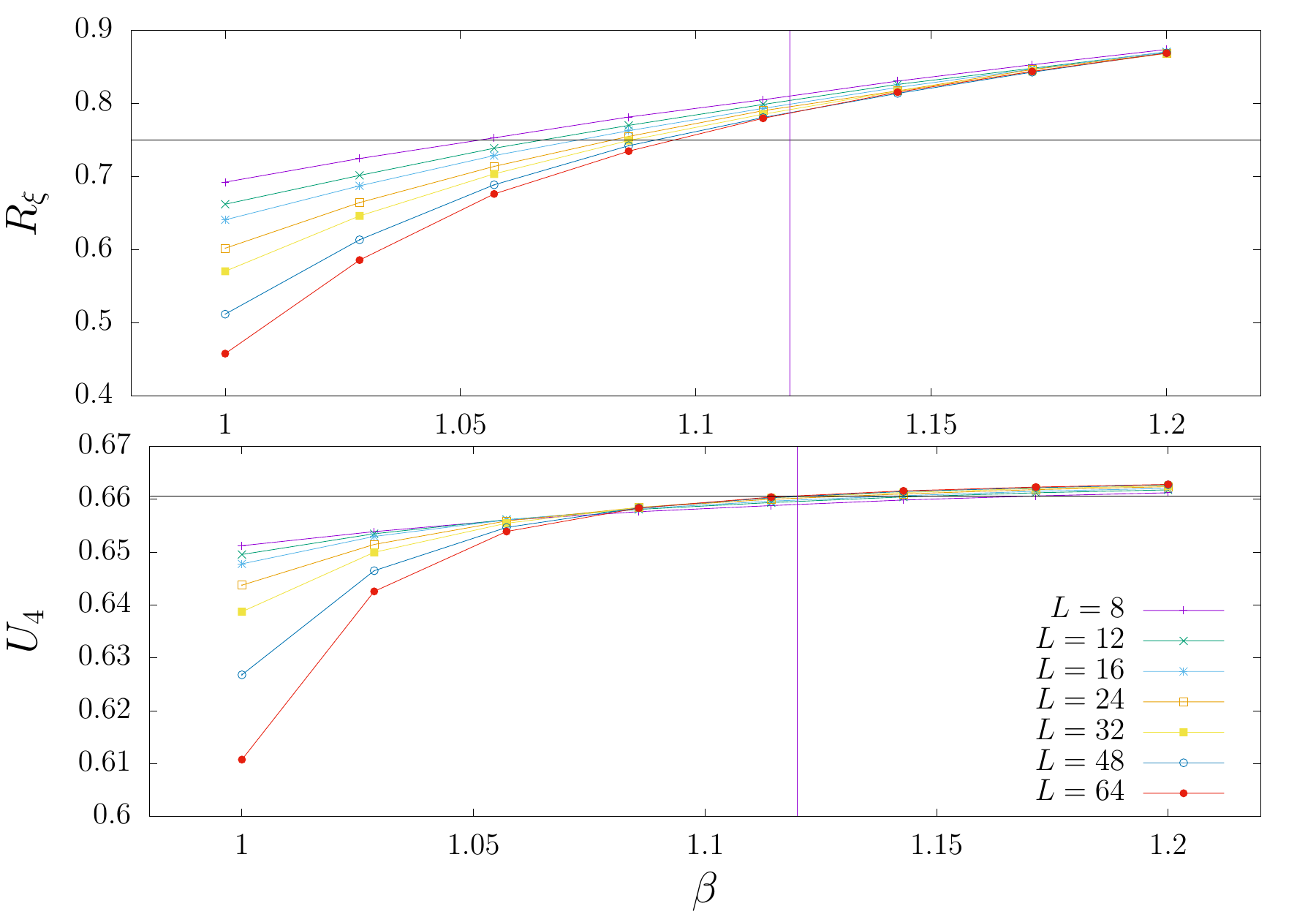}
\caption{(color online) $R_\xi$ and $U_4$ cumulants versus inverse
  temperature for the two dimensional $O(2)$ pure model on different
  lattice sizes. We have also plotted $\beta_c=1.1199(1)$,
  $R_\xi=0.7506912$ and $U_4=0.660603(12)$.}
\label{fig:XILXY}
\end{figure}

We have analyzed this model by studying the $R_\xi$ and $U_4$ curves
(see Fig. \ref{fig:XILXY}) using the fixed coupling method and
neglecting scaling corrections. We report the results in Tables
\ref{tab:extrapolationFCXYU4} and \ref{tab:extrapolationFCXYR}.  One
can see from Fig. \ref{fig:XILXY} that the correlation length data
suffer from larger corrections to scaling than the Binder cumulant
ones.  Notice that the $U_4$ analysis provided with a value of the
$\eta$ exponent compatible with the analytical one
($\eta=1/4$). Moreover, the $\nu$ exponents take very large values,
eventually diverging in the thermodynamic limit, showing the fact that
different curves of $R_\xi$ and $U_4$ are merging in the low-temperature phase, which presents QLRO behavior.

\begin{table*}[hbt!]
\caption{Independent extrapolations of the fixed coupling method data,
  using the Binder cumulant $U_4=0.6606$, for the two dimensional XY
  model.}  \centering
\label{tab:extrapolationFCXYU4}
\begin{tabular}{|c|c|c|c|c|c|c|c|}
\hline
$U_4$ &  $\nu_\xi$ &  $\nu_{U_4}$ &$\eta$ \\
\hline
0.6606   & 3.8(2)  & 3.6(3)    & 0.246(3) \\
\hline
\end{tabular}
\end{table*}

\begin{table*}[hbt!]
\caption{Independent extrapolations of the fixed coupling method data,
  using $R_\xi=0.75$, for the two dimensional XY model.}  \centering
\label{tab:extrapolationFCXYR}
\begin{tabular}{|c|c|c|c|c|c|c|c|}
\hline
$R_\xi$ &  $\nu_\xi$ &  $\nu_{U_4}$ &$\eta$ \\
\hline
0.75   & 4.0(1)  &  7.0(6)    & 0.236(1) \\
\hline
\end{tabular}
\end{table*}

\section{The magnetization and $R_\xi$ in the three-dimensional Edwards-Anderson model}
\label{EA}

To illustrate the behavior of some magnetic observables in a model
with a spin-glass low-temperature phase, we have simulated two small
lattices of the three-dimensional Edwards-Anderson model with binary
coupling, which is a classic spin glass with zero magnetization. See
Ref. \cite{janus:13} for additional details of the model (including
Hamiltonian and observables).

 In particular, we have simulated 1600 samples of $L=6$ and 8 using
the  Metropolis and the Parallel Tempering algorithms with 20 temperatures
 (204800 Monte Carlo steps per temperature) around the infinite volume
 critical inverse temperature: $\beta_c=0.908(2)$ \cite{janus:13}.

\begin{figure}[ht]
\centering
\includegraphics[width=\columnwidth, angle=0]{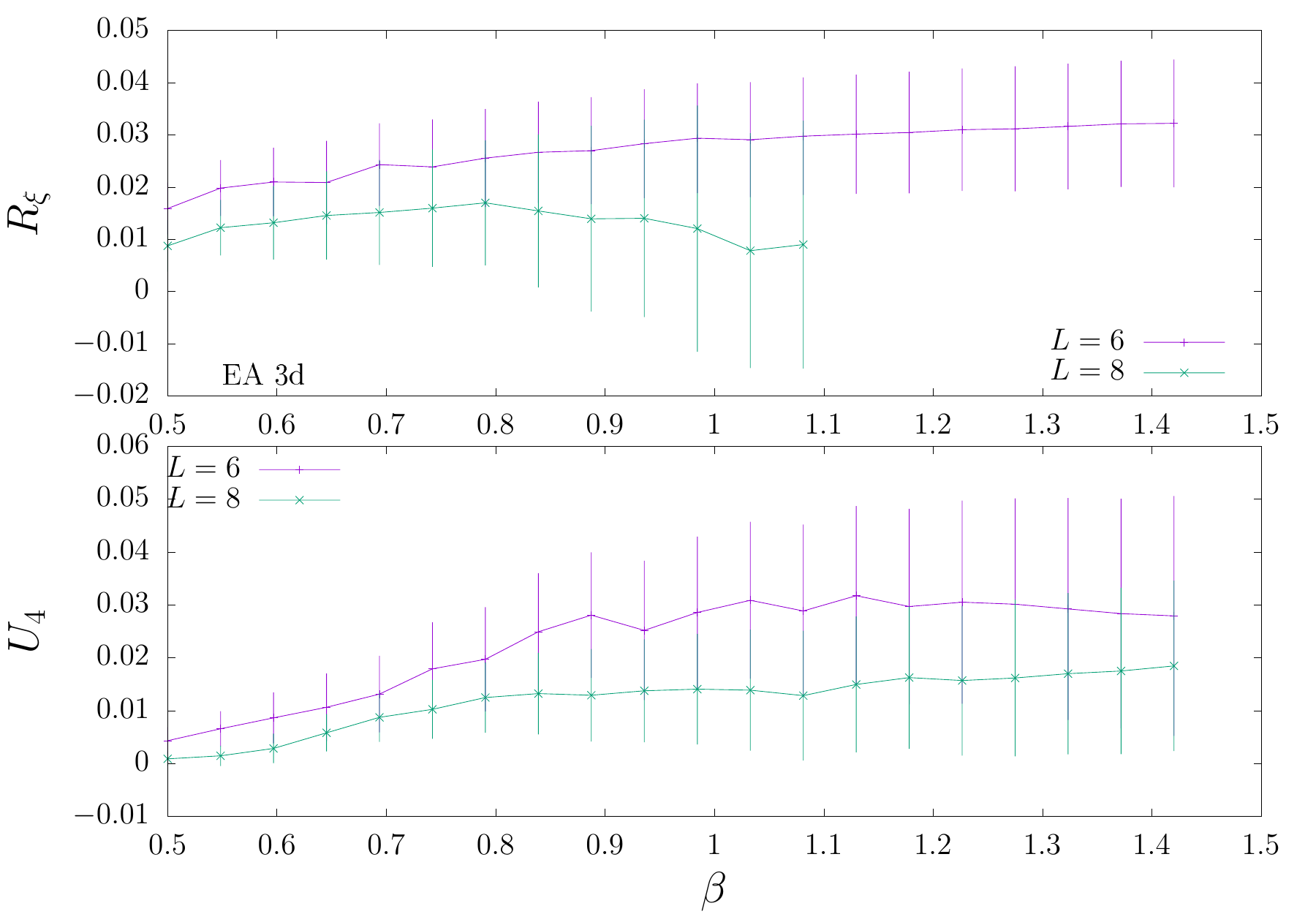}
\caption{(color online) Behavior of the correlation length in units of
  $R_\xi$the lattice size $R_\xi$ (top) and the Binder cumulant $U_4$
  (bottom) in the magnetic channel for the three dimensional
  Edwards-Anderson model. For $L=6$ and $L=8$ both observables are
  almost compatible with zero. Notice that in the low-temperature
  region we have been unable to compute $R_\xi$, even simulating a
  huge number of samples, due a signal-noise problem (the final value
  is compatible with zero and the computation of $R_\xi$ involves a
  square root).}
\label{fig:EAM}
\end{figure}

Figure \ref{fig:EAM} shows the lack of criticality of this magnetization
channel: $R_\xi$ and $U_4$ are already compatible with zero, as 
should be.

\begin{figure}[ht]
\centering
\includegraphics[width=\columnwidth, angle=0]{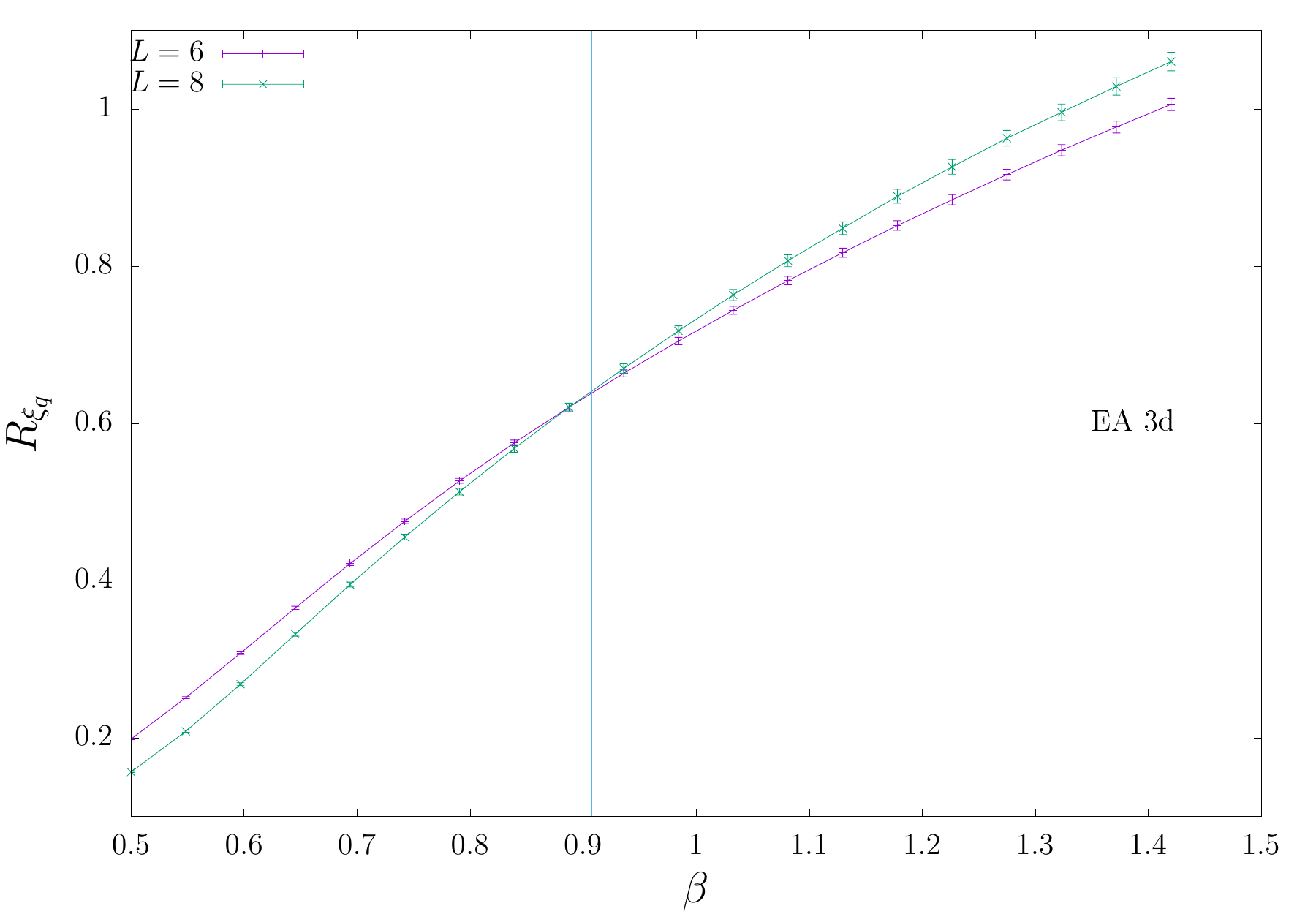}
\caption{(color online) Behavior of the overlap correlation length in
  units of the lattice size $R_{\xi_q}$ versus the inverse temperature
  for the three-dimensional Edwards-Anderson model for small lattice
  sizes. Notice that both curves cross, marking the apparent critical
  temperature of the model. The vertical line marks the infinite
  volume critical inverse temperature $\beta_c=0.908(2)$
  \cite{janus:13}.}
\label{fig:EAQ}
\end{figure}

However, the overlap channel clearly shows the PM-SG phase transition,
see Fig. \ref{fig:EAQ}. In this figure, the curves for the cumulant
$R_{\xi_q}$ cross for the two different lattice sizes.

\section{The three-dimensional $O(3)$ model}
\label{O3}

In this appendix we present our results on the three-dimensional
$O(3)$ model using the quotient method with the same methodology as
applied throughout the paper for the RAM.

We have run for this model $L=6, 12, 16, 24, 32$ and 48 lattices,
using parallel tempering with eight temperatures in the fixed range
$[0.686,0.700]$.

\begin{figure}[hbt!]
\centering
\includegraphics[width=\columnwidth, angle=0]{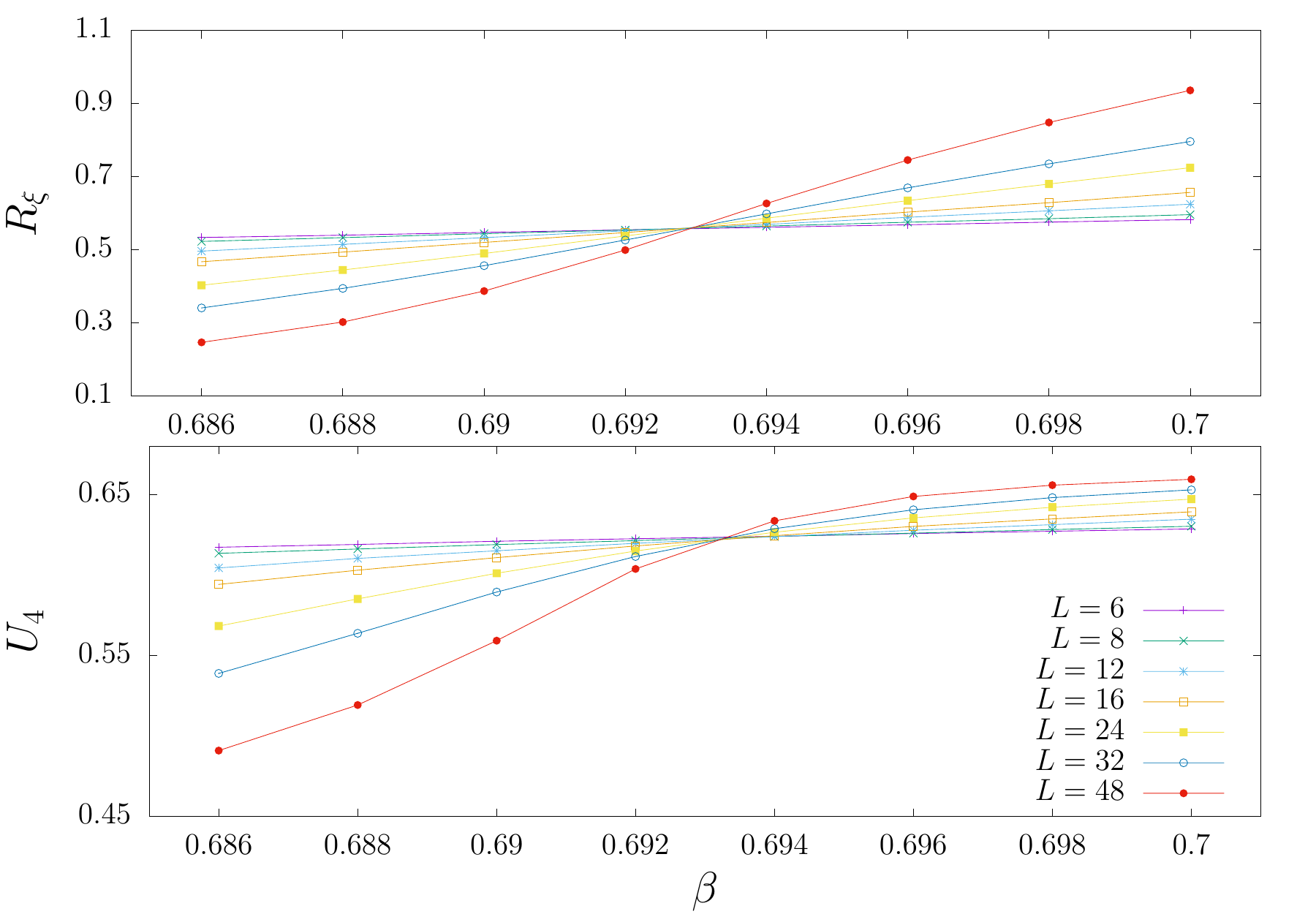}
\caption{(color online) $R_\xi$ (top) and $U_4$ (bottom) cumulants against inverse temperature for the three-dimensional $O(3)$ pure model for several lattice sizes.}
\label{fig:XILO3}
\end{figure}

In Tables \ref{tab:resultsPureXi} and \ref{tab:resultsPureB} we show
the results of the quotient method for the crossing points of $R_\xi$
and the Binder cumulant respectively (see also
Fig. \ref{fig:XILO3}). First, we have computed $\omega$ by fitting the
last column to Eq. \eqref{eq:dimensionless-quotients}. Once we have
got this value, we perform all the fits of the rest of the columns using
this $\omega$ value but for the inverse critical temperature (in this
case, the correction is $1/\nu+\omega$, and we have left completely
free this exponent in the fit). In the last row of these tables we
have reported our extrapolations to infinite volume. The agreement,
despite the small lattice sizes simulated, with the values reported in
Table \ref{tab:Models}, is pretty good. In addition, the inverse
critical temperature is fully compatible with the one reported in
Ref.~\cite{ballesteros:96} ($\beta_c= 0.69300(1)$).

\begin{table*}[hbt!]
\caption{Quotient method results for the pure $O(3)$ model ($D=0$)
  from the crossing points of $R_\xi$ for lattice sizes $L_1$ and
  $L_2$. The numbers in round brackets are the statistical error of
  the fit and the number inside the square ones are systematic errors
  originating from the uncertainty in the $\omega$ exponent used in
  the fit.}  \centering
\label{tab:resultsPureXi}
\begin{tabular}{|c|c|c|c|c|c|c|}
\hline
$L_1/L_2$  & $\beta_\text{cross}$ & $R_\xi$ & $\nu_\xi$ & $\nu_{U_4}$ &
  $\eta$ & $Q_{U_4}$\\
\hline\hline
6/12 & 0.692577(7)    & 0.5567(5) & 0.727(3) &0.697(7) & 0.024(2)  & 0.9966(3)\\
8/16 & 0.69283(3)     & 0.5587(3) & 0.718(3) &0.699(5) & 0.029(2)& 0.9972(2)\\
12/24& 0.69290(2)     & 0.5597(5) & 0.716(3) &0.705(7) & 0.034(2)& 0.9982(2)\\
16/32& 0.69293(2)   & 0.5601(4) & 0.714(4) &0.706(7) & 0.037(2)& 0.9986(2)\\
24/48& 0.69298(1)   & 0.5617(4) & 0.717(6) &0.72(1)  & 0.036(3)  & 0.9990(3)\\ \hline \hline
$\infty$ & 0.6930(3)  & 0.5630(5)[6] & 0.708(4)[2] & 0.717(9)[2] & 0.043(2)[2] & $\omega=0.92(13)$\\
\hline \hline
\end{tabular}
\end{table*}

\begin{table*}[hbt!]
\caption{Quotient method results for pure $O(3)$ model from the
  crossing points of $U_4$ for lattice sizes $L_1$ and $L_2$. The
  numbers between round brackets are the statistical error of the fit
  and the number inside the square ones are systematic errors
  originating from the uncertainty in the $\omega$ exponent used in
  the fit.}  \centering
\label{tab:resultsPureB}
\begin{tabular}{|c|c|c|c|c|c|c|}
\hline
$L_1/L_2$  & $\beta_\text{cross}$ & $U_4$ & $\nu_\xi$ & $\nu_{U_4}$ &
  $\eta$ & $Q_{U_4}$ \\
\hline\hline

6/12  & 0.6942(1)    &0.6243(2) &0.731(8) &0.728(3) &-0.015(3) & 1.0168(7) \\
8/16  & 0.69368(5)   &0.6236(1) &0.720(6) &0.718(2) &-0.003(2) & 1.0130(6) \\
12/24 & 0.69320(3)   &0.6223(2) &0.721(8) &0.716(3) & 0.017(2) & 1.0083(8) \\
16/32 & 0.69309(2)   &0.6217(1) &0.720(8) &0.714(4) & 0.021(2) & 1.0064(8) \\
24/48 & 0.69305(2)   &0.6214(1) &0.72(1)&0.716(6) & 0.025(3) & 1.005(1) \\ \hline \hline
$\infty$ & 0.69300(3)& 0.6202(2)[2] &0.707(3)[1] & 0.72(1)[1] & 0.042(3)[3] & $\omega=0.98(9)$\\

\hline \hline
\end{tabular}
\end{table*}

\end{document}